\documentclass[hyper,letterpaper]{JHEP3} 

\pdfoutput=1 

\usepackage{epsfig,multicol,bbm}

\newcommand{\reef}[1]{(\ref{#1})}

\title{Quarks in an  External Electric Field in Finite Temperature Large $N$ Gauge Theory}

\author{{Tameem Albash, Veselin Filev, Clifford V. Johnson, Arnab Kundu}\\
	Department of Physics and Astronomy\\ University of Southern California\\ Los Angeles, CA 90089\\
	E-mail: \email{albash@usc.edu}, \email{filev@usc.edu}, \email{johnson1@usc.edu}, \email{akundu@usc.edu}}

\preprint{\arXivid{0709.1554}}	

\abstract{We use a ten dimensional dual string background to study aspects of the physics of large $N$ four dimensional $SU(N)$ gauge theory, where its fundamental quarks are charged under a background electric field.  The theory is ${\cal N}=2$ supersymmetric for vanishing temperature and electric field.  We work in a limit where the quarks do not back--react.  At zero temperature, we observe that the electric field induces a phase transition associated with the dissociation of the mesons into their constituent quarks. This is an analogue of an insulator--metal transition, since the system goes from being an insulator with zero current (in the applied field) to a conductor with free charge carriers (the quarks). At finite temperature this phenomenon persists, and the dissociation transition becomes subsumed into the more familiar meson melting transition. Here, the dissociation phenomenon reduces the critical melting temperature.}


\begin{document} 


\section{Introduction}
\label{sec:introduction}
\indent In the series of studies of the dynamics of (non--backreacting) quarks\cite{Karch:2002sh} in large $N$ $SU(N)$ gauge theory using holographically dual\cite{Susskind:1998dq} geometries in string theory\cite{Maldacena:1997re,Gubser:1998bc,Witten:1998qj,Witten:1998zw}, the non--perturbative phenomena that can occur in the presence of external electromagnetic fields have recently become recognized as readily amenable to computation. The case of external magnetic field at zero temperature was studied in refs.\cite{Filev:2007gb,Filev:2007qu}, where a number of interesting phenomena (such as spontaneous symmetry breaking) were readily extracted. There was also a recent study of quarks at non--zero baryon number density (see refs.\cite{Karch:2007pd,O'Bannon:2007in}) in the presence of a background electric field, where properties (such as the conductivity) of the system were uncovered, again using holographic techniques. Very recently, the finite temperature case of external magnetic field was studied in some detail, with the uncovering of an interesting phase diagram arising from the competing effects of temperature and magnetic field \cite{Albash:2007bk}.

In this paper, we study the same theory ($SU(N)$ gauge theory with non--backreacting hypermultiplet quark flavours\cite{Karch:2002sh}) ---an ${\cal N}=2$ supersymmetric theory in its ground state--- in the presence of a background electric field, at both zero and finite temperature\footnote{We note that another group will present results in this area in a paper to appear shortly\cite{Erdmenger:2007bn}.}.  Right at the outset, we might expect very different phase structure than the magnetic case since the electric field and the finite temperature are working together. The quarks and anti--quarks that are bound together to make mesons are oppositely charged under the electric field, and so its presence should serve to reduce that binding energy, making it easier for finite temperature to shake the mesons apart. This is indeed what we see. Moreover, we can quite readily identify a sharp first order phase transition even at zero temperature corresponding to the dissociation of the mesons into their
constituent quarks above a critical value of the electric field.  Furthermore, since the mesons are electrically neutral, the dissociation represents a rather clean example of a ``metal--insulator'' phase transition, where charge carriers are
liberated, creating a current in response to the applied field.

The paper is organised as follows. In section two we introduce the dual string background we are going to use throughout the paper and the D7--brane probes that represent the quarks. By switching on gauge fields of a particular sort on the world--volume, we can introduce a background electric field. This same gauge field may also source a non--trivial current in the gauge theory, and we allow for this possibility in our computations. In the next section we solve the equations of motion that result from our apparatus, and observe the aforementioned phase structure. In particular, when the dissociated or melted phase is present, the system has a non--zero result for the response current, representing the flow of the liberated quarks, the charge carriers. We map out the phase structure of the system in this section. There is an important subtlety that we observe. At high enough values of the electric field, some of our solutions for the embeddings of the D7--branes have conical singularities that we can characterize, but do not fully understand. It is a possibility that these solutions are corrected in the full string theory.  However, we believe that their ultimate fate does not change the phase diagram we present.  Until we know conclusively the fate of these solutions, we are careful to clearly show how they arise and where they lie in the phase diagram which we uncover. It is to be expected that future work in this area may yield techniques that can help further refine our embeddings.  In section four we observe some aspects of the meson spectrum of the system, and we briefly conclude in section five.
%
\section{The String Background} \label{sec:string}
%
We consider the following form for the metric of the AdS$_5$--Schwarzschild$\times S^5$ background:
\begin{eqnarray}
ds^2 /\alpha' &=& \frac{1}{4 r^2 R^2} \left(-\frac{f^2}{\tilde{f}} dt^2 + \tilde{f} d \vec{x}^2\right) + \frac{R^2}{r^2}  dr^2 + R^2 \cos^2 \theta d \Omega_3^2 \nonumber \\
&& \hskip2cm + R^2 d \theta^2 + R^2 \sin^2 \theta d \phi^2 \ , \\  
f  &=& 4r^4 - b^4 \ ,\quad  
\tilde{f} = 4r^4 + b^4 \nonumber \ . 
\end{eqnarray}
The AdS$_5$ time and space coordinates are given by $t\equiv x^0$ and $\vec{x}\equiv(x^1,x^2,x^3)$ respectively (which the dual gauge theory will also have) and $r\in[0,\infty)$, which is the AdS radial coordinate, which we will sometimes denote as $x^4$.  We use standard polar coordinates on the $S^5$, given by:
\begin{eqnarray}
d\Omega_5^2&=&d\theta^2+\cos^2\theta d\Omega_3^2+\sin^2\theta d\phi^2\nonumber \ ,\\ \mathrm{and}\quad
d\Omega_{3}^2&=&d\psi^2+\cos^2\psi d\beta+\sin^2\theta d\gamma^2 \ .
\end{eqnarray}
The scale $R$ determines the gauge theory 't Hooft coupling according to $R^2=\alpha^\prime\sqrt{g_{\rm YM}^2 N}$. The parameter $b$ sets the radius of the horizon of the black hole, which in turn sets the temperature, according to $T=b/\pi R^2$. The coordinate $r$ is related to the traditional radial coordinate $u$ by the coordinate transformation $r^2 = \frac{1}{2} \left(u^2 + \sqrt{u^4 - b^4} \right)$.
To proceed, we consider the action for the probe D7--brane to second order in $\alpha'$:
\begin{eqnarray}
\frac{S_{\mathrm{D7}}}{N_f}&=& S_{\mathrm{DBI}} + S_{\mathrm{WZ}} \nonumber \\
&=& - T_{\mathrm{D7}} \int d^8 \xi \ \mathrm{det}^{1/2} \left(P\left[ G_{a b} \right] + P \left[B_{a b} \right] + 2 \pi \alpha' F_{a b} \right) \nonumber\\ &&\hskip2cm + \left(2 \pi \alpha' \right)^2  \frac{\mu_7}{2} \int F_{(2)} \wedge F_{(2)} \wedge P\left[ C_{(4)} \right] \ ,
\end{eqnarray}
where $P\left[ G_{a b} \right]$, $P\left[ B_{a b} \right]$, and $P\left[ C_{(4)} \right]$ are the pullback of the background metric, the $B$--field, and the 4--form potential (sourced by the $N$ D3--branes) onto the D7--brane worldvolume respectively, $F$ is the field strength on the D7--brane worldvolume, and $ T_\mathrm{D7} =\mu_7 / g_s $.  In order to introduce our background electric field (and current), we consider an ansatz for the world--volume gauge field of the form~\cite{Karch:2007pd}:
\begin{equation} 
A_1(r)  =- E t + B(r) \ .
\end{equation}
This ensures a constant electric field $F_{01}=E$ in the gauge theory; $B(r)$ sources a current $J^1$ on the world--volume \cite{Karch:2007pd}.  For the D7--brane embedding, we consider  an ansatz of the form:
\begin{equation} 
\theta \equiv  \theta(r) \ .
\end{equation}
The resulting effective action for the D7--brane is:
\begin{eqnarray}
S_\mathrm{D7}&=&- T_{\mathrm{D7}} N_f  \int d^8 \xi \ \frac{\cos^3 \theta(r)}{16 r^5} \left[ \tilde{f}^2 f^2 \left(1+ r^2 \theta'(r)^2\right) \right. \\
&&\hskip 4cm  \left. - 4 \left(2 \pi \alpha' \right)^2 r^4 \left \{ - f^2 \tilde{f} B'(r)^2 + 4\tilde{f}^2 R^4 E^2 \left(1+ r^2 \theta'(r)^2\right)   \right\} \right]^{1/2} \ . \nonumber 
\end{eqnarray}
It is convenient to define dimensionless quantities ${\tilde r},
{\tilde B}$ and ${\tilde E}$ {\it via}:
\begin{equation} 
r = b \tilde{r}  \ ,  \  \ B(r) = \frac{b}{2 \pi \alpha'} \tilde{B}(\tilde{r}) \ ,  \   \ E = \frac{b^2}{2 \pi \alpha' R^2} \tilde{E} \ ,  \  \  \theta(r) = \theta(\tilde{r}) \nonumber \ ,
\end{equation}
which reduces the action to:
\begin{eqnarray}
S_\mathrm{D7}&=& - 2 \pi^2 V N_f T_{\mathrm{D7}} b^4 \int d t \int d \tilde{r} \ \tilde{\mathcal{L}} \ , \label{eqt:action} \\
\mathrm{where}\qquad \tilde{\mathcal{L}} &=& \frac{\cos^3 \theta(\tilde{r})}{16 \tilde{r}^5} \left[ \tilde{g
}^2\left(g^2 - 16 \tilde{r}^4 \tilde{E}^2 \right) \left(1 + \tilde{r}^2 \theta'(\tilde{r})^{2} \right) +  4 \tilde{r}^4  g^2 \tilde{g} \tilde{B}'(\tilde{r})^{2} \right]^{1/2} \nonumber \ , \\ \mathrm{with}\qquad 
g &=& 4 \tilde{r}^4 - 1 \ , \quad \tilde{g} = 4 \tilde{r}^4 +1 \ . \nonumber
\end{eqnarray}
The equations of motion derived from equation \reef{eqt:action} for $\tilde{B}(r)$ introduces a constant of motion~$\tilde{T}$.  We can invert the equation of motion to write a first order differential  equation for $\tilde{B}(r)$ in terms of  $\tilde{T}$: 
%
\begin{eqnarray}
\tilde{B}'(\tilde{r}) &=& - \tilde{T}  \frac{4 \tilde{r} \tilde{g} \sqrt{\left(g^2 - 16 \tilde{r}^4 \tilde{E}^2 \right)\left(1 + \tilde{r}^2 \theta'(\tilde{r})^2\right)}}{g \sqrt{\tilde{g} \left( 64 \tilde{r}^6 \left(\tilde{S}^2 g^2 - \tilde{T}^2\tilde{g}^2 \right) + g^2 \tilde{g}^3 \cos^6 \theta(\tilde{r}) \right)}} \ .
\end{eqnarray}
In the limit of $\tilde{r} \to \infty$, the asymptotic solution for ${\tilde B}(r)$ is given by:
\begin{eqnarray}
\lim_{\tilde{r} \to \infty} \tilde{B} (\tilde{r}) & = & \tilde{b} + \frac{\tilde{T}}{2 \tilde{r}^2} + \dots \nonumber
\end{eqnarray}
Normalisability of the solution requires us to take $\tilde{b} = 0$.  The constant $\tilde{T}$ is related to the (vacuum expectation value) vev of the current by the following relation \cite{Karch:2007pd}:
\begin{equation}\label{eqt:current}
\langle J^1 \rangle = \langle \bar{\psi} \gamma^1 \psi \rangle = - 4 \pi^3 \alpha' b^3 V N_f T_{\mathrm{D7}}  \tilde{T}\ ,
\end{equation}
where $V$ is the spatial volume of the gauge theory.  It is convenient to exchange the fields $\tilde{B}(r)$ for the constant $\tilde{T}$ by performing a Legendre transformation, defining a new action $\tilde{I}_{D7}$:
\begin{eqnarray}
I_\mathrm{D7} &=& S_\mathrm{D7}  - \int d^8 \xi \  F_{41} \frac{\delta S_{D7}}{\delta F_{41}} \nonumber \\
&=& -2 \pi^2 V N_f T_{\mathrm{D7}} b^4 \int d t \int d \tilde{r} \ \left( \tilde{\mathcal{L}} - \tilde{B}' (\tilde{r}) \frac{\partial \tilde{\mathcal{L}}}{\partial \tilde{B}' (\tilde{r}) } \right) \nonumber \\
&=& -2 \pi^2 V N_f T_{\mathrm{D7}} b^4 \int d t \int d \tilde{r} \left[ \frac{\sqrt{\left(1 + \tilde{r}^2 \theta'(\tilde{r})^2\right)}}{16 \tilde{r}^5 g \sqrt{\tilde{g}}}  \right. \nonumber \\
&&\hskip5cm \left. \times  \sqrt{\left(g^2 - 16 \tilde{r}^4 \tilde{E}^2 \right) \left( -64 \tilde{r}^6  \tilde{T}^2\tilde{g}^2  + g^2 \tilde{g}^3 \cos^6 \theta(\tilde{r}) \right)} \right] \nonumber \\
&=& -2 \pi^2 V N_f T_{\mathrm{D7}} b^4 \tilde{I}_{D7}\ ,\label{eqt:legendre}
\end{eqnarray}
where $F_{41} = \partial_r A_1$.  It is a simple check to show that ${\delta \tilde{I}_{\mathrm{D7}}}/{\delta \tilde{T}} = \tilde{B}'(\tilde{r})$.  The expression under the square root in equation \reef{eqt:legendre} must be positive in order to keep the action real.  Either both terms should be negative or both should be positive, and they must change sign at the same radial distance~$\tilde{r}_\ast$.  This results in two conditions:
\begin{equation} \label{eqt:cond1}
\left(4 \tilde{r}_\ast^4 -1 \right)^2 - 16 \tilde{r}_\ast^4 \tilde{E}^2  =  0 \ ,
\end{equation}
\begin{equation} \label{eqt:cond2}
-64 \tilde{r}_\ast^6  \tilde{T}^2 \left(4 \tilde{r}_\ast^4 +1 \right)^2 + \left( 4 \tilde{r}_\ast^4 -1 \right)^2 \left(4 \tilde{r}_\ast^4 +1 \right)^3 \cos^6 \theta(\tilde{r}_\ast)  = 0 \ .
\end{equation}
Equation \reef{eqt:cond1} determines $\tilde{r}_\ast$ in terms of
$\tilde{E}$:
\begin{equation}
\tilde{r}_\ast^2 = \frac{\tilde{E} + \sqrt{\tilde{E}^2 + 1}}{2} \ ,
\end{equation}
and equation \reef{eqt:cond2} yields $\tilde{T}$ in terms of  $ \tilde{r}_\ast$, and $\theta(\tilde{r}_\ast)$:
\begin{equation} \label{eqt:ohm}
\tilde{T}^2 =\frac{\left( 4 \tilde{r}_\ast^4 -1 \right)^2 \left(4 \tilde{r}_\ast^4 +1 \right)^3 \cos^6 \theta(\tilde{r}_\ast)}{64 \tilde{r}_\ast^6 \left(4 \tilde{r}_\ast^4 +1 \right)^2} \ .
\label{tsolution}
\end{equation}
Note that equation \reef{eqt:ohm} relates the current to the electric field in a form of Ohm's law, and this was used in ref.~\cite{Karch:2007pd} to determine the conductivity of the fundamental matter in the quark--gluon plasma.  We refer to $\tilde{r}_\ast$ as the ``vanishing locus''.  From the action ${\tilde I}_{D7}$ defined in equation \reef{eqt:legendre}, we can derive the equation of motion for the embeddings $\theta(\tilde{r})$. Its exact form is not illuminating, and so we do not display it here. In the limit of large $\tilde{r}$, the equation of motion asymptotes to:
\begin{equation}
\frac{d}{d \tilde{r}} \left( \tilde{r}^5 \theta' (\tilde{r}) \right) + 3 \tilde{r}^3 \theta(\tilde{r}) = 0 \ ,
\end{equation}
which has as a solution:
\begin{equation}
\theta(\tilde{r}) = \frac{\tilde{m}}{\tilde{r}} + \frac{\tilde{c}}{\tilde{r}^3} \ .
\end{equation}
The constants $\tilde{m}$ and $\tilde{c}$ are related to the bare quark mass and the condensate respectively, in a manner that is by now very standard \cite{Karch:2002sh,Kruczenski:2003be,Babington:2003vm,Albash:2006ew}.  In our notation,
the exact relationship is given by:
\begin{eqnarray}
m_q &=& \frac{b \  \tilde{m}}{2 \pi \alpha'} \ , \qquad
\langle \bar{\psi} \psi \rangle = - 8 \pi^3  \alpha' V N_f T_{D7}  b^3 \tilde{c} \ .
\end{eqnarray}
%
\section{Properties of the Solutions}
\subsection{Exact Results at Large Mass}
%
It is instructive to study the properties of the quark condensate as a function of the bare quark mass for large mass. This corresponds to $\tilde{m}\gg \tilde{E}$ in terms of the dimensionless quantities introduced in the previous section.  This limit constrains us to consider only the so--called ``Minkowski'' embeddings for the probe D7--brane for which the vev of the current (defined in equation (\ref{eqt:current})) vanishes by virtue of equation (\ref{tsolution}).  To extract the behaviour of quark mass and condensate we linearize the equation of motion obtained from equation (\ref{eqt:legendre}) in the same way as described in ref.~\cite{Albash:2007bk}. Rewriting the result in terms of dimensionful parameters, we find the following analytic behaviour for the condensate:
\begin{eqnarray}
\langle \bar{\psi}\psi \rangle\propto -c=\frac{R^4\hat{E}^2}{4m}+\frac{b^8+4b^4R^4 \hat{E}^2+8R^8 \hat{E}^4}{96m^5}+O\left(\frac{1}{m^7}\right) \ ,
\end{eqnarray}
where $\hat{E} = 2 \pi \alpha' E$.  This may suggest that for high enough bare quark mass, the condensate vanishes.  However unlike the results for an external magnetic field~\cite{Albash:2007bk}, the condensate approaches zero on the positive side of the condensate axis.  We verify these results in the next section using numerical techniques.
%
\subsection{The Case of Vanishing Temperature}
%
It is instructive to first study the zero temperature case, since a number of important features appear in this case.  For this simple case, the energy scale is set by $R\sqrt{\hat{E}}$, and therefore it is convenient to introduce the dimensionless quantities ${\hat r}$ and ${\hat m}$ {\it via}:
\begin{equation}
r=R\sqrt{E}\, {\hat r}\ ;\quad m=R\sqrt{E}\,{\hat m}\ .
\label{dimless}
\end{equation}
We solve the equation of motion for $\theta(\hat{r})$ using a shooting technique by imposing infrared boundary conditions described in refs.~\cite{Albash:2006ew,Albash:2006bs,Karch:2006bv}. For ``Minkowski'' embeddings, in order to avoid a conical singularity at $\theta = \pi/2$ \cite{Karch:2006bv}, we impose the following boundary condition:
\begin{equation}\label{eqt:minkow}
 \theta'(\hat r)|_{\theta = \pi/2}= -\infty \ .
\end{equation}
For embeddings reaching the pseudo-horizon we find it numerically convenient to shoot forward (towards infinity) and backward (towards the origin) from the vanishing locus ${\hat r}_\ast=1$.  The boundary condition ensuring the smoothness of solutions across the vanishing locus can be determined from the equation of motion itself, and it is found to be:
\begin{eqnarray}
&& \theta'(\hat{r}_\ast)= \frac{\partial \hat u}{\partial \hat r}\frac{\partial\theta}{\partial \hat u} ;~~~\theta_0 \equiv \theta(\hat{r}_\ast);\\
&&\frac{\partial\theta}{\partial\hat u}=-\left.\frac{1}{\hat{u}}\tan\frac{{\theta}_0}{2}\right|_{\hat u(\hat r_\ast)}.\nonumber
\end{eqnarray}
From the above boundary condition we can see that the embeddings reaching the vanishing locus at $\theta=\pi/2$ avoids the conical singularity by having a diverging derivative.  Next we proceed to discuss our numerical findings.

Numerical analysis shows that there are three different classes of embeddings, which can be classified by the topology of the probe D7--brane.  First, we have the smooth ``Minkowski'' embeddings that close above the vanishing locus, due to the shrinking of the $S^3$ that the probe brane wraps.  Second, we have the embeddings that reach the vanishing locus before the $S^3$ shrinks.  These embeddings can in turn be classified into two different categories. The first are singular solutions that cross the vanishing locus and close before reaching the origin. These have a conical singularity at the closing point because the derivative does not diverge.  We discuss these solutions further below. The second are smooth solutions that pass through the vanishing locus and reach the origin with no singular behaviour. These different embeddings are  summarized in figures \ref{fig:embed0} and \ref{fig:em02}.
\FIGURE{\epsfig{file=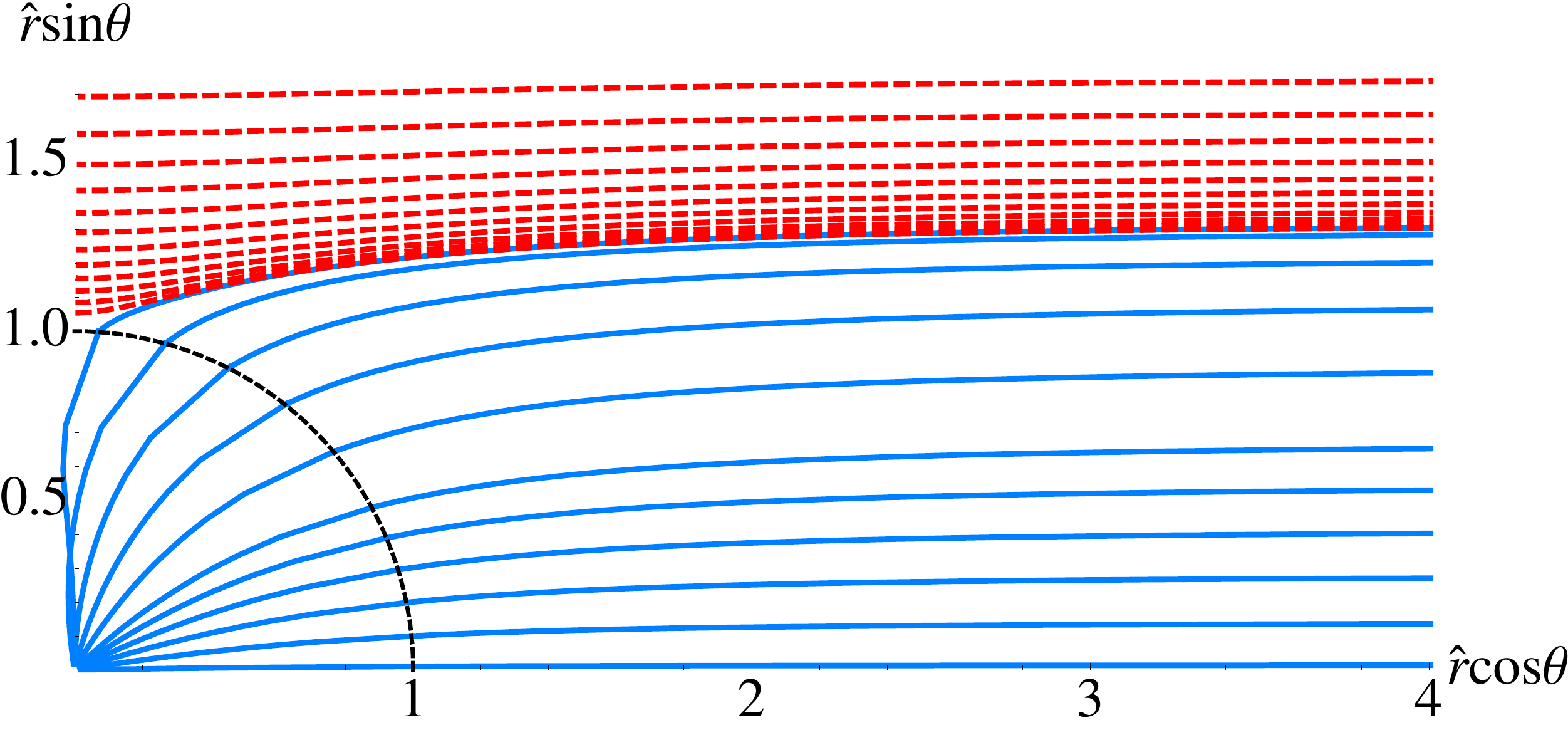,width=9cm}
 \caption{\small The dashed red curves represent solutions smoothly closing before reaching the vanishing locus.  The solid blue remaining curves correspond to embeddings passing the vanishing locus (denoted by a semi--circular dashed black curve).}
 \label{fig:embed0}}
\FIGURE{\epsfig{file=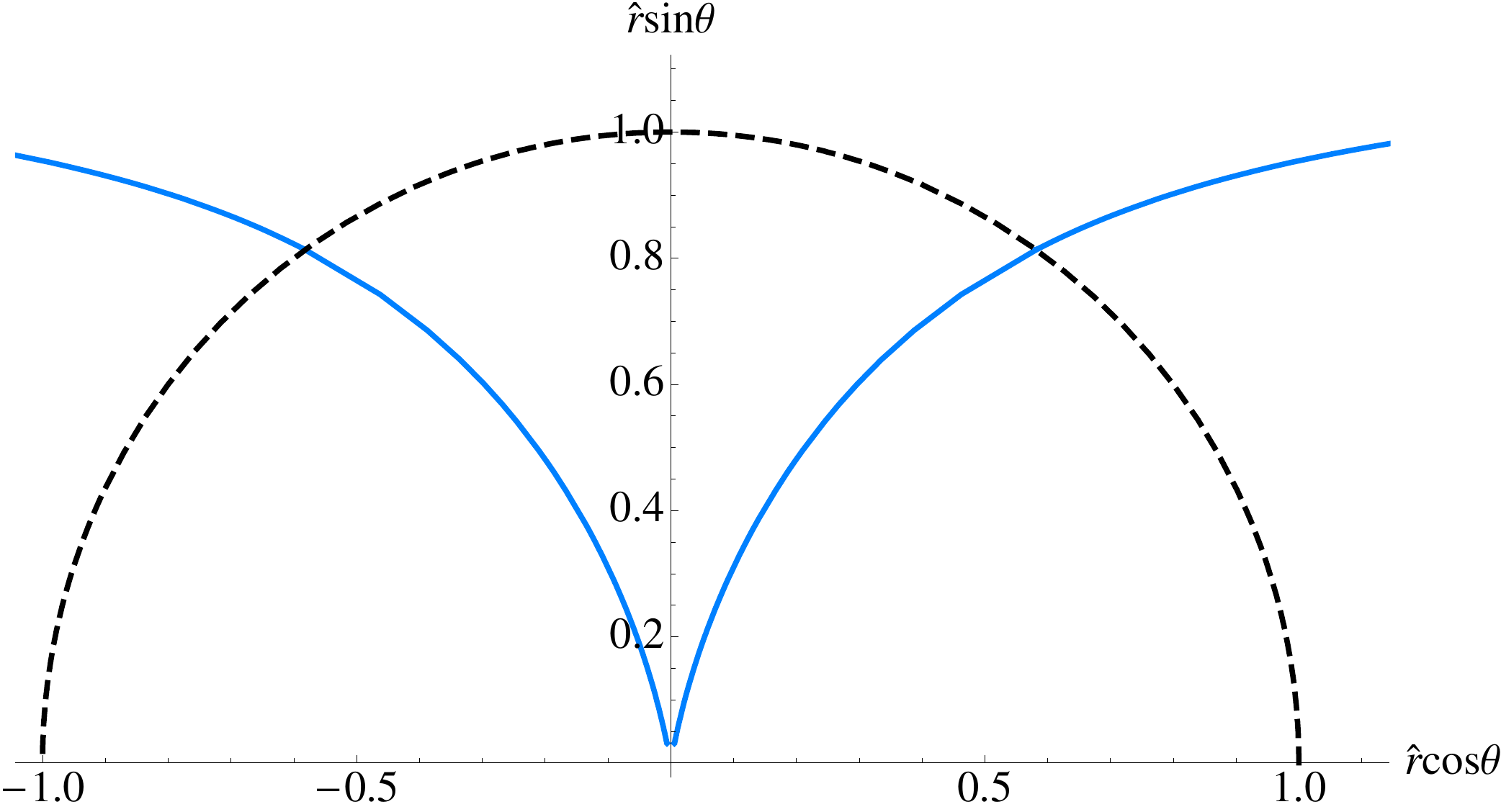,width=2.8in}\epsfig{file=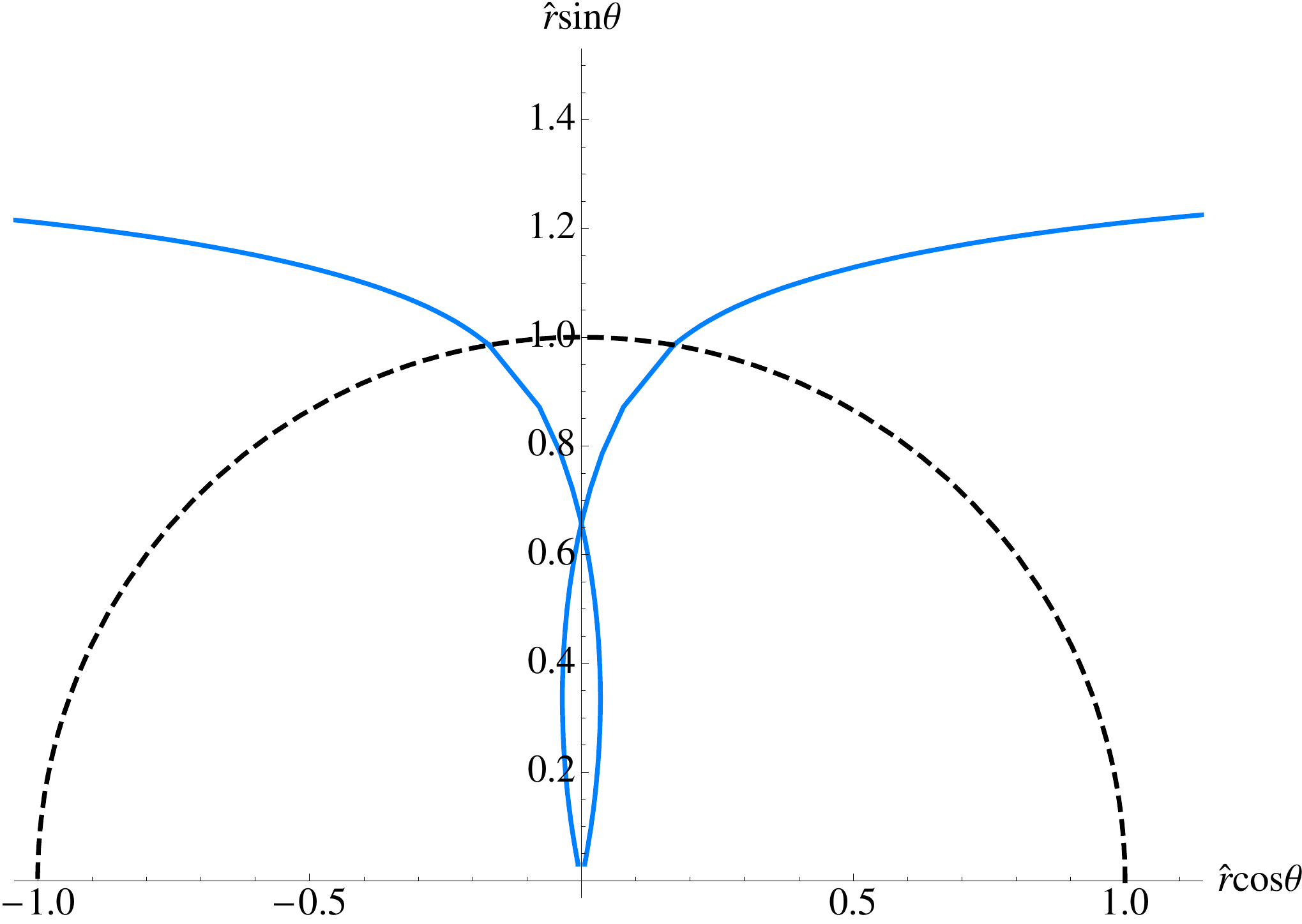,width=2.8in}
\caption{\small Two kinds of embeddings that pass the vanishing locus corresponding to D7--brane closing at the origin (left), and forming a conical singularity (right) respectively. The semi--circular dashed (black) curve corresponds to the vanishing locus.}
\label{fig:em02}}
We can see explicitly the conical singularity appearing for the brane (second curve in figure \ref{fig:em02}) since it intersects
the vertical axis before reaching the event horizon.  This class of singular solutions starts appearing after a certain value of
$\theta_0 = \left(\theta_{0}\right)_{\rm min}$ that depends on the magnitude of the electric field; this value is close to $\pi/2$ (which corresponds to the critical embedding), and the singular solution persists until $\theta_0=\pi/2$.
 
The existence of these embeddings suggests the possibility of a transition in topology of the probe brane as a function of the
bare quark mass, as is well known to exist at finite temperature.  Here, the external electric field alone induces the transition. The Minkowski embeddings simply have a shrinking $S^3$, while the embeddings reaching the origin are distinguished by having an $S^3$ shrinking as well as touching the AdS horizon. However the presence of singular solutions makes the nature of the transition subtle, as we discuss later. We later see that the finite temperature case also shares a similar classification of embeddings.
  
From the asymptotic behaviour of these embeddings we can extract the condensate as a function of the bare quark mass. In figure
\ref{fig:cm0} we show this dependence. There are two important mass scales. One is where the phase transition between the two types of embedding occurs, which we label ${\hat m}_{\rm cr}$, and the other is where the singular solutions first appear, which we label ${\hat  m}^\ast$.  These  values are in turn determined by the parameters $(\theta_0)_{\rm cr}$ and $(\theta_0)_{\rm min}$, respectively.
\FIGURE{\epsfig{file=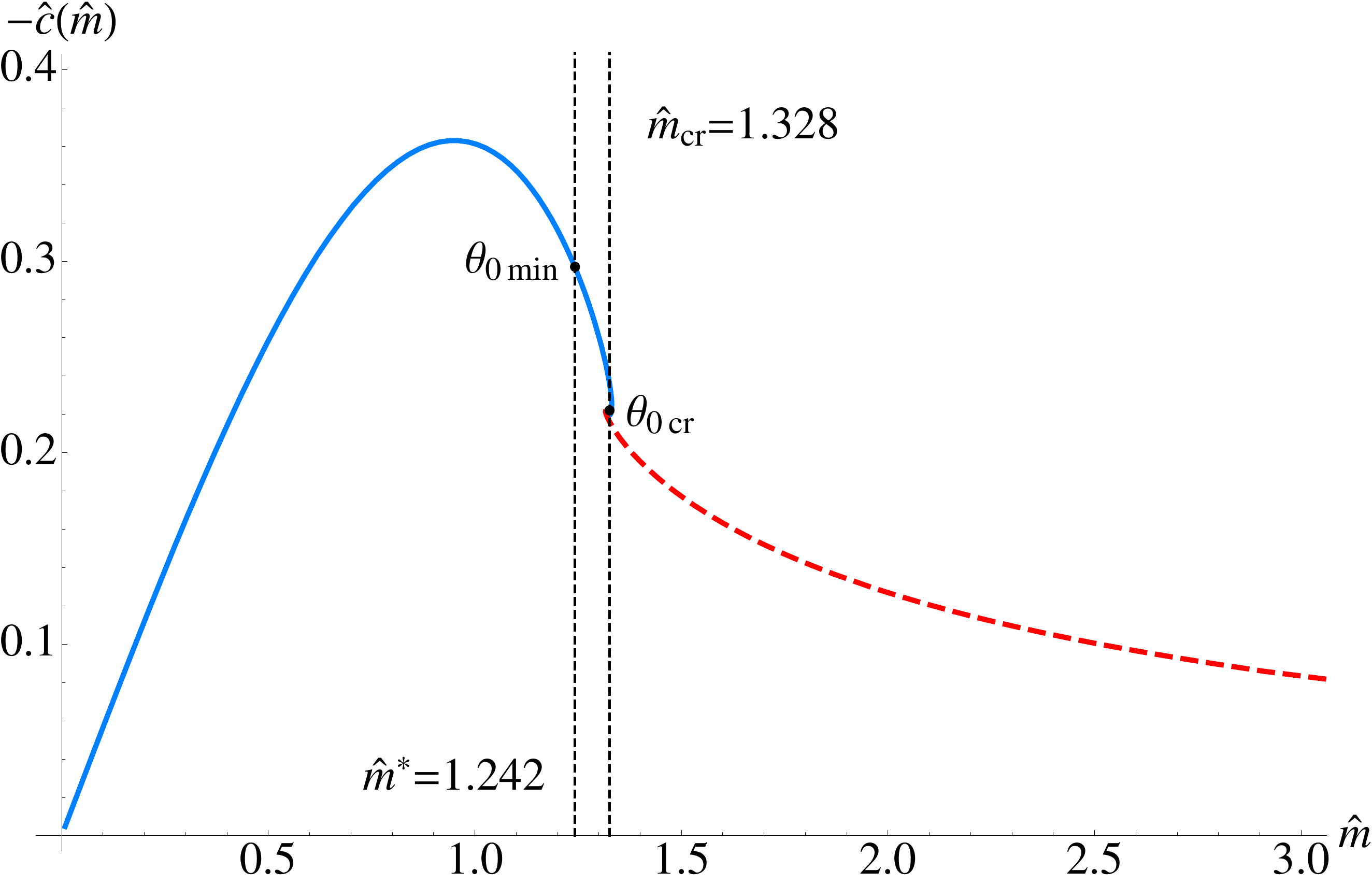,width=8cm}\epsfig{file=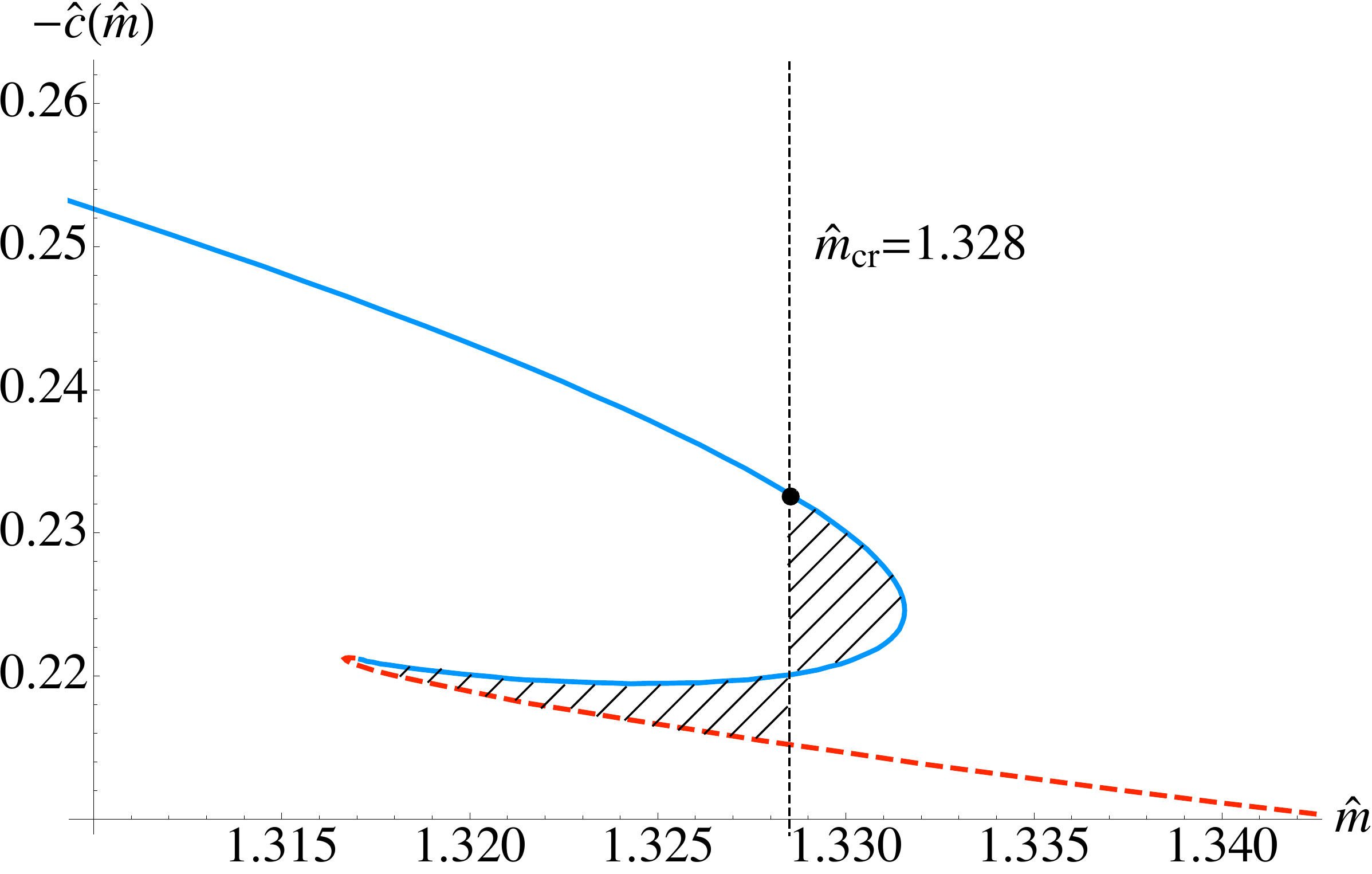,width=8cm}
\caption{\small The condensate as a function of bare quark mass ${\hat m}$. The curve segment coming in from the left (solid blue)  corresponds to embeddings reaching the vanishing locus, and the curve segments going out to the right (dashed red) correspond to Minkowski embeddings. There is a family of  solutions that contain a conical singularity between the vanishing locus and the origin. They exist for $(\theta_0)_{\rm min}<{\theta}_0<(\theta_0)_{\rm cr}$. On the right is a magnification of the turn--around region where these segments join, showing multi--valuedness.  Since it has been a good guide in earlier work \cite{Albash:2006ew}, we use an equal--area law argument to determine the presence of a first order phase transition at ${\hat m}_{\rm cr}$ where there is a jump from one type of embedding to another.}
 \label{fig:cm0}}

In figure \ref{fig:cm0}, the multi--valuedness is indicative of a phase transition, as was speculated earlier based on the general arguments.  Furthermore we observe that there is no chiral symmetry breaking, since in all cases the condensate vanishes for vanishing mass. This is to be contrasted with the case of having an external magnetic field where chiral symmetry breaking is observed~\cite{Albash:2007bk}.

The latter observation fits our intuition that the mesons in this theory have a binding energy that is proportional to the constituent quark mass~\cite{Kruczenski:2003be}. For a given quark mass, the introduction of an electric field reduces the binding energy of the quarks. This causes two things to happen: First, it inhibits the formation of a chiral condensate, and second, it ultimately dissociates the mesons into its constituent quarks.  This dissociation is in fact a transition from an insulating to a conducting phase, mediated by the external electric field. On the dual gravity side this corresponds to a transition from Minkowski
embeddings to embeddings that reach the origin.

We must note that the appearance of the singular solutions (those that have a conical singularity) are not well understood at the moment.  There is therefore the possibility that there is as yet not identified intermediate phase right after the mesons dissociate.  
We can make our dissociation transition explicit. From equation~(\ref{dimless}) we can deduce the exact dependence of $m^\ast$ and $m_{\rm cr}$ on the electric field $\hat{E}$ as $m^\ast=R\sqrt{\hat{E}}\, {\hat m}^\ast$ and $m_{\rm cr}=R\sqrt{\hat{E}}\, {\hat  m}_{\rm cr}$.  The latter is the value of the critical mass for a given electric field.  Using an equal--area law (see ref.~\cite{Albash:2006ew})\footnote{It would be interesting to explore comparing the results of the equal--area prescription with the results of the thermodynamic potential computation using an appropriate brane probe action (DBI supplemented with terms corresponding to the response to the external fields.). This we will leave to a future project.}, we can also independently (as a test of our methods for later) determine this dependence numerically for our solutions, and we display this in figure~\ref{fig:phase0}.  We also include the critical mass for which the singular solutions appear as a function of the field. We see that the analytic behaviour deduced above is nicely confirmed by the numerics.
\FIGURE{\epsfig{file=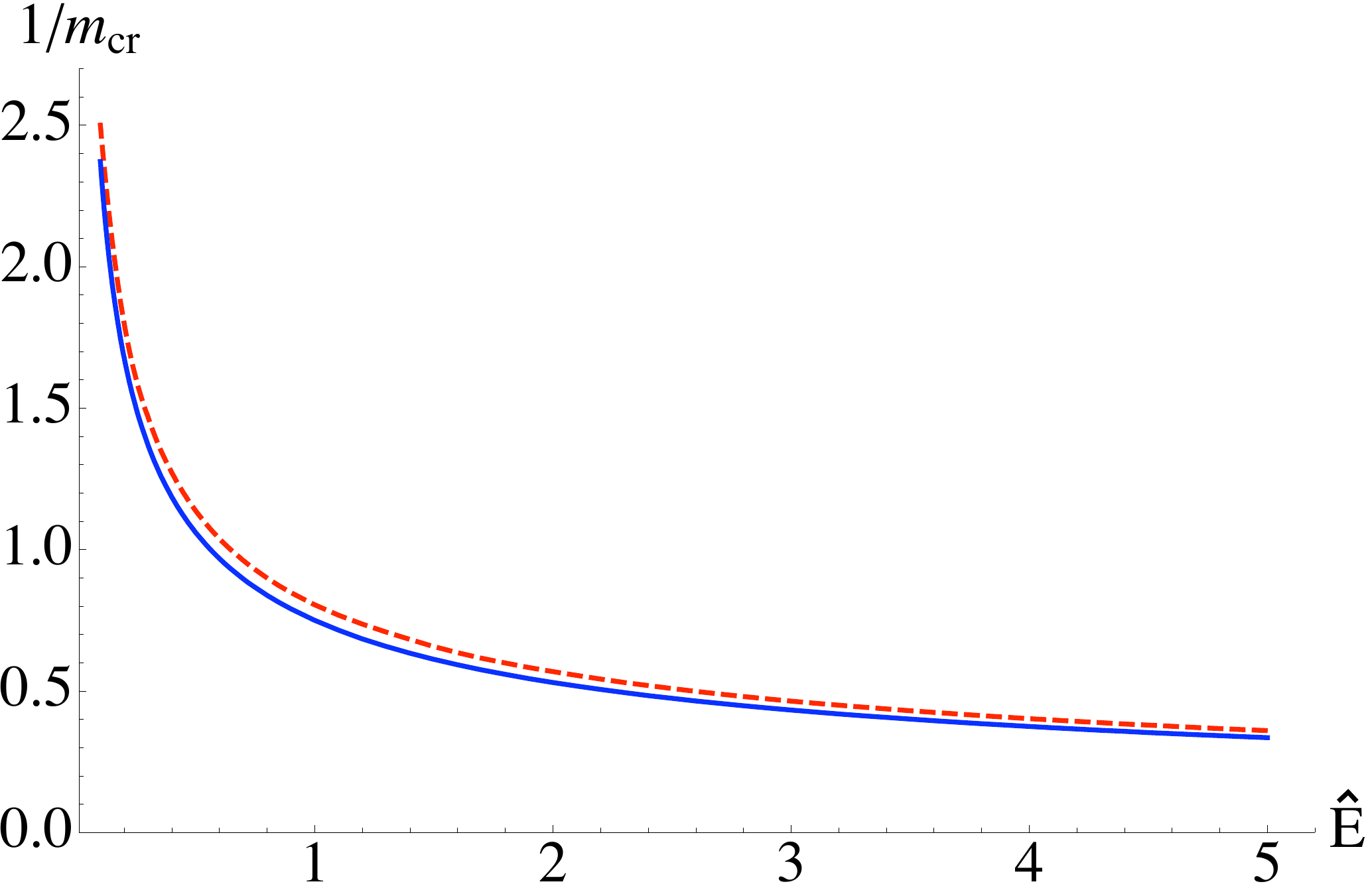,width=9cm}
\caption{\small The phase diagram showing the electric field induced first order phase transition separating the dissociated phase from the phase with  stable mesons. The blue (solid) curve separates the  phases. The region within the blue (solid) curve and red (dashed) curve represents the region where the embeddings with a conical singularity become thermodynamically relevant.  Note that $m_{\rm cr}$ has units of $R$ and $\hat{E}$ is dimensionless.}
\label{fig:phase0}}
Indeed the singular solutions seem to be unavoidable in the phase diagram (at least at the level of our analysis). However they seem to lie very close to the dissociation transition and, in the limits ${\hat E}\to 0$ and ${\hat E}\to\infty$, the two curves tend to
merge.

Presently we do not completely understand the role played by the singular solutions in the physics. One simple possibility (but not the only one) is that stringy corrections smooth out the singularity in the interior, while preserving the asymptotic behaviour. This would mean that we have a simpler phase diagram than that in figure~\ref{fig:phase0} ({\it i.e.}, with no dashed line).
In fact, a study of the meson spectrum appears to hint at the fact that the gauge theory is actually ``blind'' to the physics below the vanishing locus, at least for the type of questions we are asking.  We elaborate on this in section \ref{section:meson}.

We next turn to the finite temperature case. It is worth noting that a simple check of our results in the next section is that the large electric field (compared to temperature) limit, should recover the physics that we have seen here, and indeed it does.
%
\subsection{The Case of Finite Temperature}
%
In order to proceed, we again solve the equation of motion for $\theta(\tilde{r})$ numerically using Mathematica. We use a similar shooting technique and boundary conditions outlined in the previous section. For Minkowski embeddings we impose equation (\ref{eqt:minkow}).  For black hole embeddings, the appropriate initial condition to ensure smoothness of the embedding when we stitch the solutions across $\tilde{r}_\ast$ is:
\begin{eqnarray}
&& \theta'(\tilde{r}_\ast)= \frac{\partial \tilde u}{\partial \tilde r}\frac{\partial\theta}{\partial \tilde u} \ , \quad \theta_0\equiv\theta(\tilde{r}_\ast) \ , \\
&&\frac{\partial\theta}{\partial\tilde u}=\frac{(3\tilde u^4-1)-\sqrt{(3\tilde u^4-1)^2+9\tilde u^4(\tilde u^4-1)\tan^2\theta_0}}{3\tilde u(\tilde u^4-1)\tan\theta_0}|_{\tilde u(\tilde r_\ast)}\nonumber \ .
\end{eqnarray}
We show several solutions for $\tilde{E}=1$ in figure \ref{fig:embeddings}.
\FIGURE{\epsfig{file= 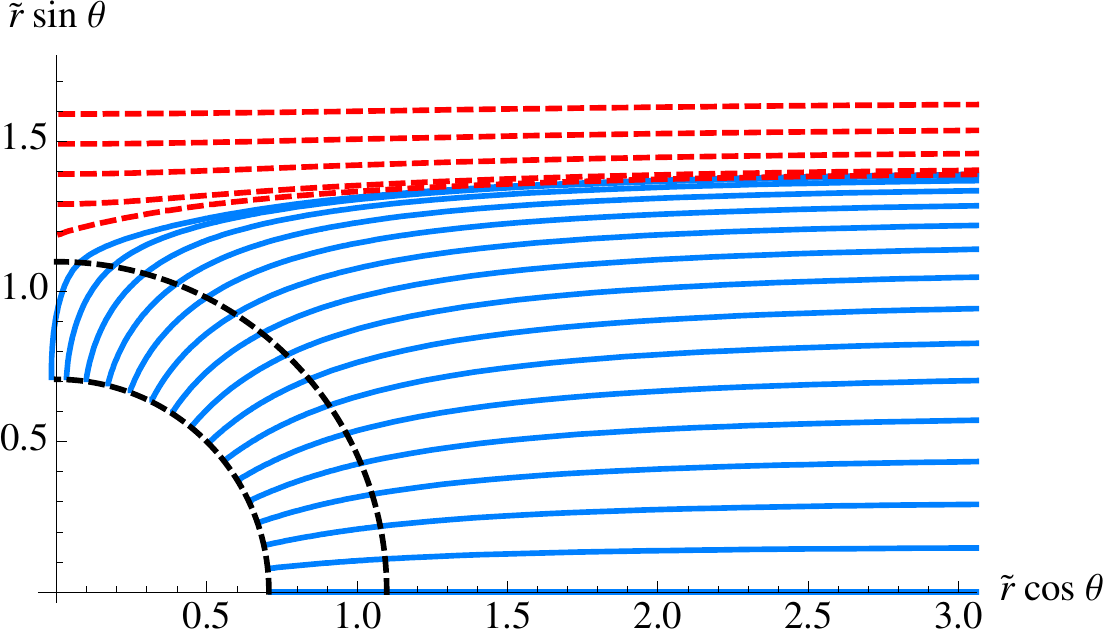,width=4in}
  \caption{\small Several solutions for classical D7--brane embeddings with $\tilde{E}=1$.  The inner semi--circle corresponds to the event horizon, and the outer semi--circle corresponds to the vanishing locus.}
\label{fig:embeddings}}
We can extract from these embeddings the condensate $\tilde{c}$ as a function of the bare quark mass $\tilde{m}$ by analyzing the asymptotic behavior of the solutions.  We show several of these solutions for various dimensionless electric field values in figure \ref{fig:c vs m}.
\FIGURE{\epsfig{file= 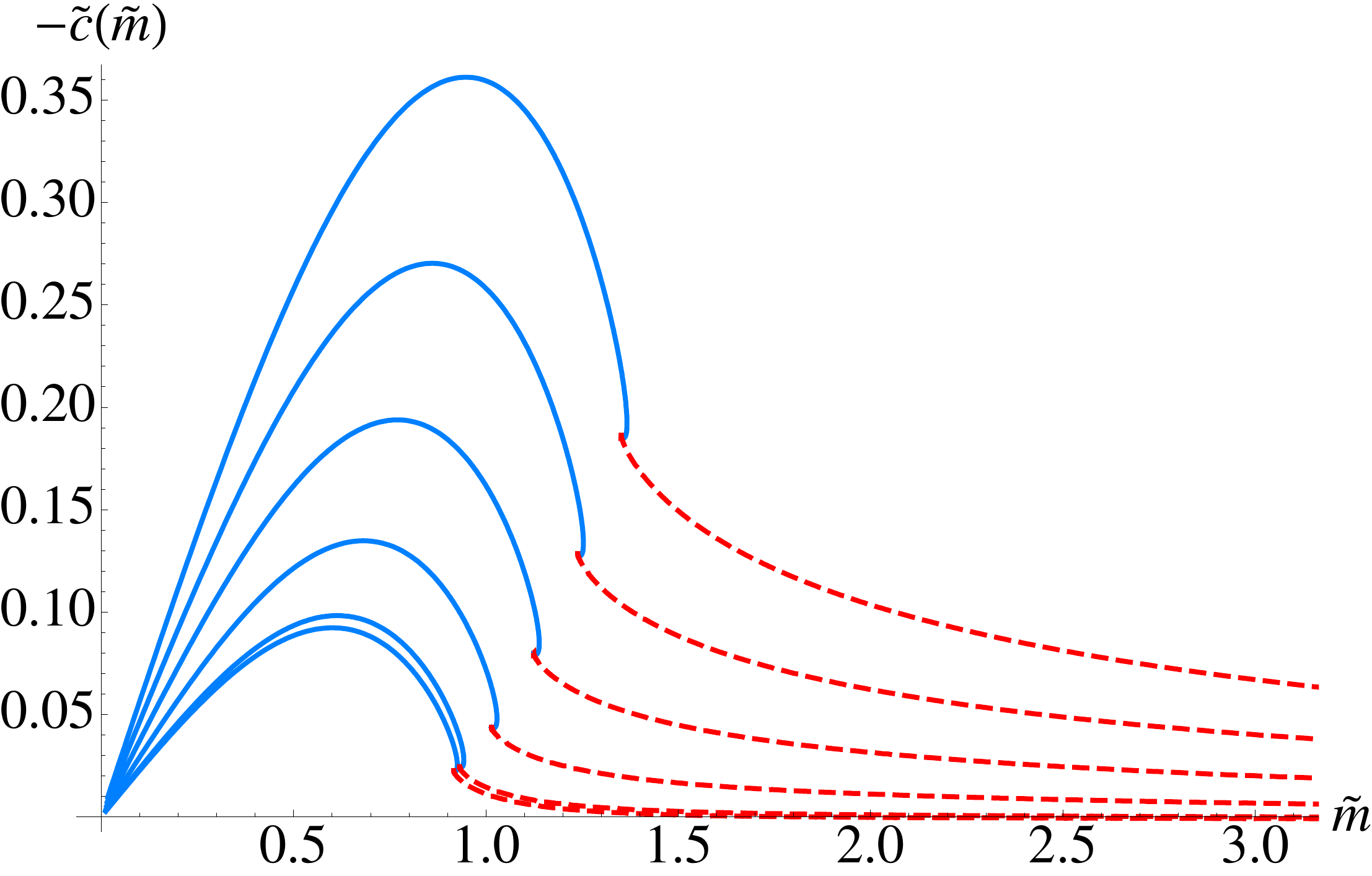,width=4in}
\caption{\small The condensate as a function of bare quark mass for increasing electric field.  From bottom to top, the electric field values are $\tilde{E} = 0.01,0.1,0.3,0.5,0.7,0.9$.}
 \label{fig:c vs m}}
The first order phase transition described in refs.~\cite{Babington:2003vm,Albash:2006ew} persists in the presence of an electric field.  We determine the critical mass at which this transition occurs using an equal--area law described in ref.~\cite{Albash:2006ew}, and we present the phase diagram in figure~\ref{fig:phase}.  The numerical results confirm our intuition: at zero temperature, the mass of the meson is proportional to the bare quark mass, and hence the binding energy is proportional to the bare quark mass~\cite{Kruczenski:2003be}.  Turning on an electric field decreases the binding energy (the electric field will pull the quark--antiquark pairs apart), and only mesons built out of sufficiently heavy quarks--{\it i.e.} above a critical mass--will survive.  At finite temperature, we can think of the increasing of the critical mass as equivalent to the decreasing of the critical temperature.  Therefore, lowering the binding energy causes mesons to melt at lower critical temperatures.  In addition, since $\tilde{c} = c / b^3$, the magnitude of the dimensionless condensate increases with decreasing critical temperature (as we see in figure~\ref{fig:c vs m}).  Furthermore, the stronger the electric field, the more initial binding energy is required and therefore a higher critical mass is needed.

We can deduce the asymptotic behavior of the phase diagram by considering large electric fields.  The dominant energy scale is set by the electric field, and therefore in that limit we are effectively at zero temperature.  To see this, one should note that the energy scale in the geometry is set by $r_\ast$, and for strong electric field, $r_\ast$ is much larger than the position of the event horizon.  Therefore, we should expect to reproduce the results for pure AdS$_5\times S^5$ in an external electric field described in the previous section.  In the case of pure AdS$_5\times S^5$ in an external electric field, the only energy scale is given by $R \sqrt{E}$, and therefore we can easily show that the bare quark mass and condensate scale as:
\begin{eqnarray}
m & \propto& R \sqrt{E} \ , \nonumber \\
c &\propto& R^3 E^{3/2} \ .
\end{eqnarray}
This result suggests that:
\begin{eqnarray}
1/\tilde{m}_\mathrm{crit} & \propto& 1/ \sqrt{\tilde{E}} \ .
\end{eqnarray}
This dependence was checked numerically (and shown in figure \ref{fig:phase}) and appear to be valid for $\tilde{E} > 2$.
\FIGURE{\epsfig{file= 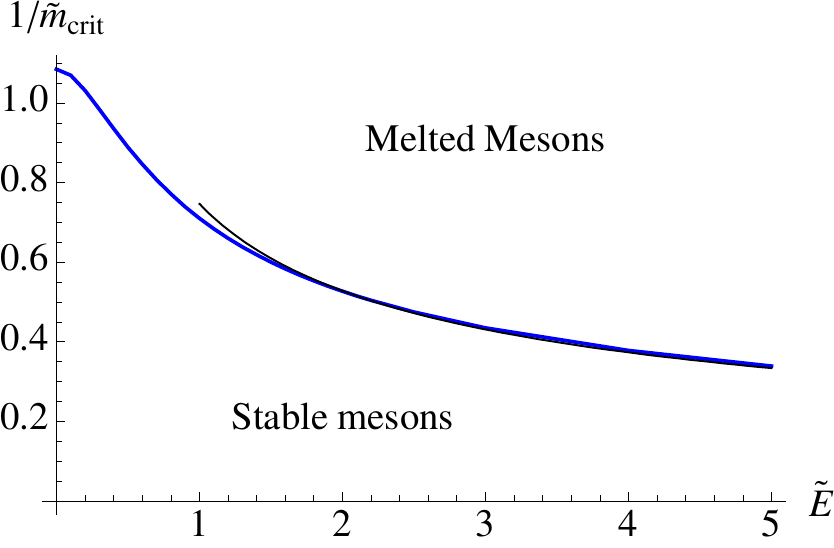,width=4in}
\caption{\small The phase diagram depicting the first order phase transition, denoted by a heavy line (blue) that separates the melted and stable meson phase.  The thin black curve is the best fit curve given by $\mathrm{const.} \times \tilde{E}^{-1/2}$ for large electric field.}
\label{fig:phase}}
\FIGURE{\epsfig{file= 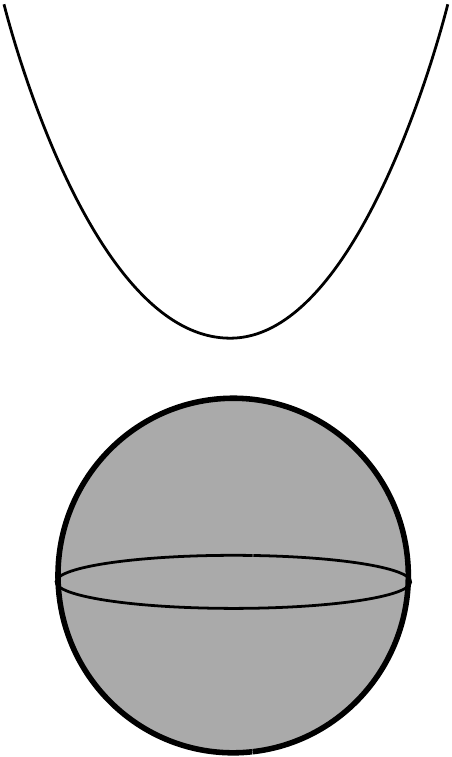,width=1.2in}\epsfig{file= 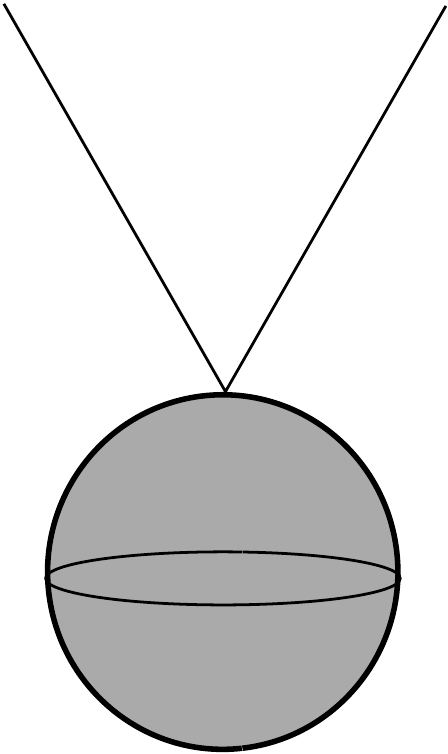,width=1.2in}\epsfig{file= 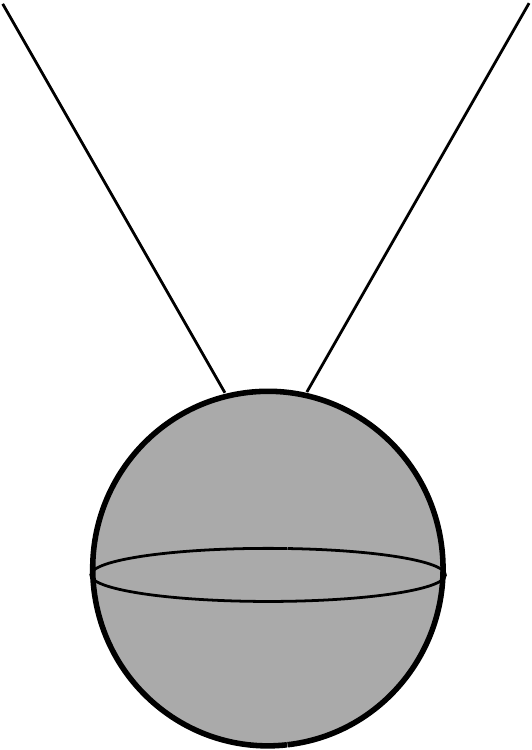,width=1.45in}
 \caption{\small \small For zero electric field and finite temperature, the Minkowski and black hole solutions are separated by a ``critical" embedding, which has conical singularity at the event horizon.}  \label{fig:A1}}
It is instructive to compare the phase transition to the one for zero electric field and finite temperature. In the latter case the Minkowski and black hole embeddings are separated by a ``critical'' embedding with a conical singularity at the event horizon as shown in figure~\ref{fig:A1} . The straightforward generalization of this picture to the case of finite electric field suggests the existence of a ``critical'' embedding with a conical singularity at the vanishing locus ($r=r_\ast$). However, similarly to the zero temperature case, a more detailed study of the transition from Minkowski to black hole embeddings reveals an interesting third class of embeddings. These are embeddings which enter the vanishing locus at $r_\ast$, and have a conical singularity above the event horizon. See figure~\ref{fig:A2}.
  \FIGURE{\epsfig{file= 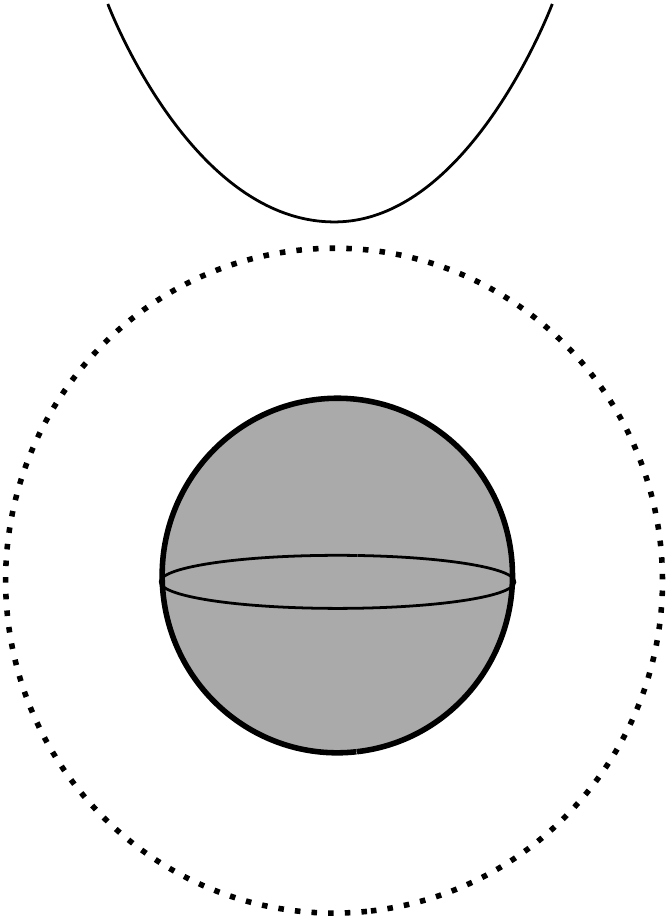,width=1.2in}\epsfig{file= 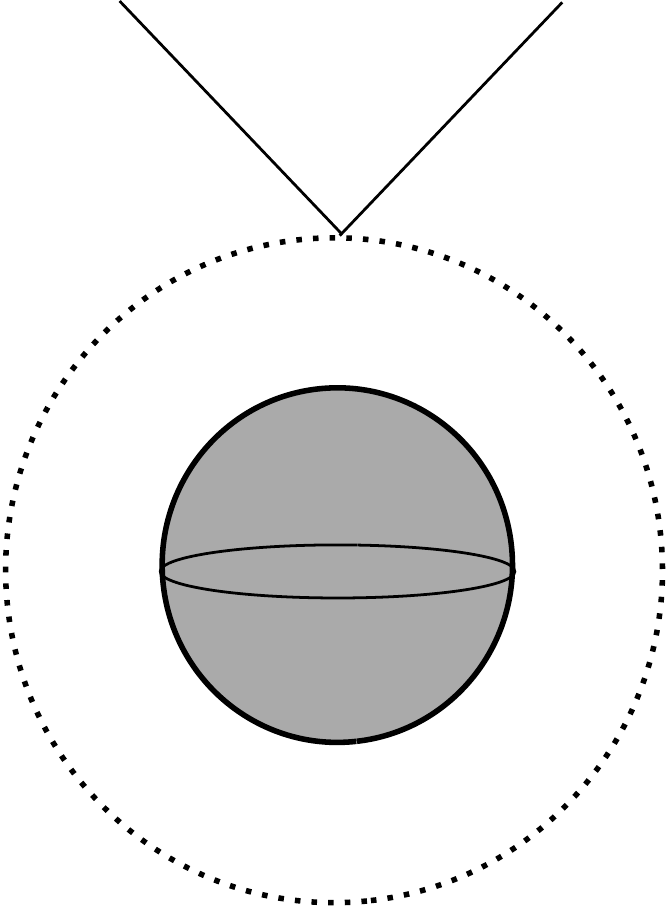,width=1.2in}\epsfig{file= 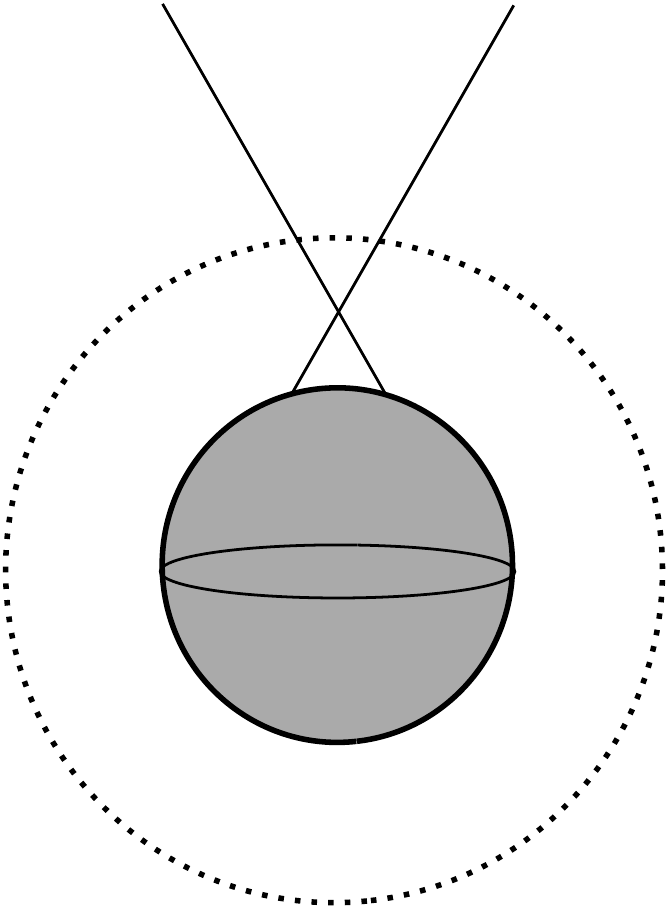,width=1.2in}\epsfig{file= 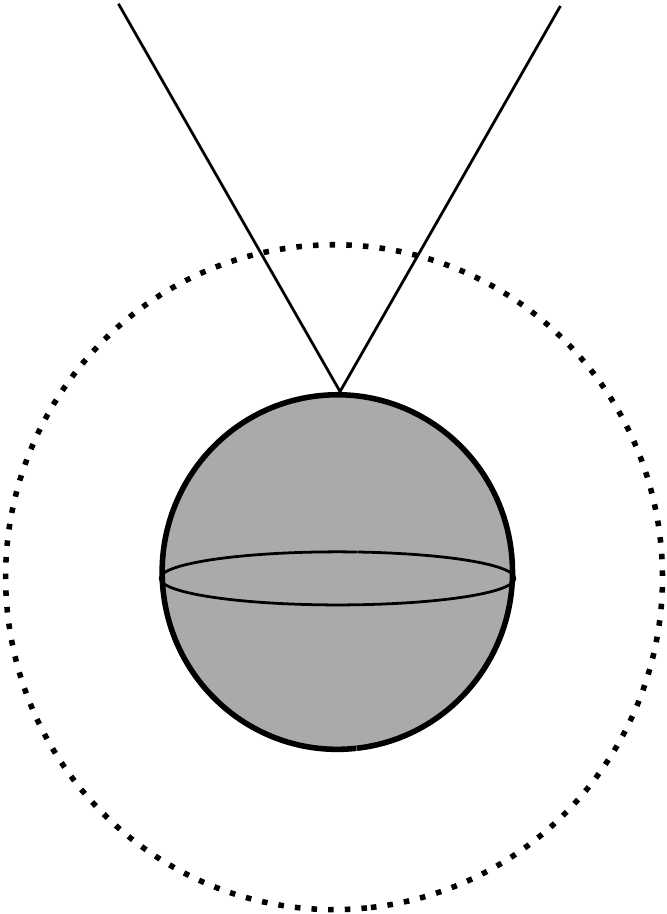,width=1.2in}\epsfig{file= 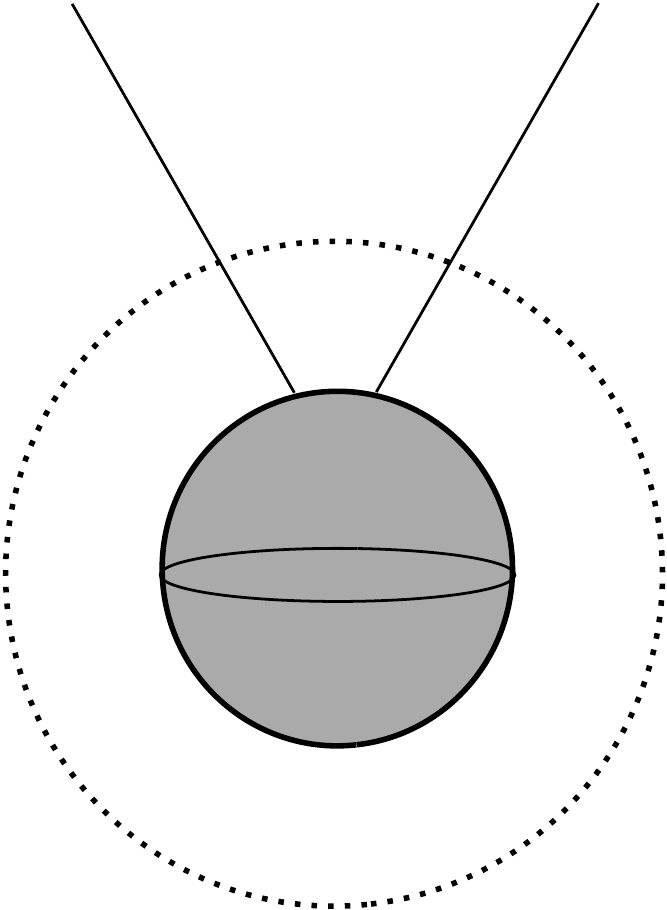,width=1.2in}
 \caption{\small \small For a finite electric field there are two ``critical" solutions corresponding to embeddings with a conical singularity at the vanishing locus ($r=r_{\ast}$) and at the event horizon respectively. The intermediate solutions correspond to black hole embeddings, which have a conical singularity between the event horizon and the vanishing locus.}
 \label{fig:A2}}
These solutions correspond to embeddings which enter the vanishing locus at $\theta_0$ close to $\pi/2$. Such solutions extend from $\theta_0 = \pi/2$ to some minimal value $(\theta_{0})_{\rm min}$, below which the embeddings have no conical singularity and simply fall into the black hole.  In figure \ref{fig:B} one can see examples of embedding for different values of $\theta_0$.  Finite temperature adds richness to the picture by adding the possibility that these singular solutions may or may not be bypassed by the otherwise melting transition induced by temperature. This depends entirely on the value of the electric field for a given temperature, and for sufficiently high electric field the singular solutions appears unavoidable (at least at this level of analysis --- as mentioned earlier, there may well be stringy corrections to the conical geometry in the interior). 
  \FIGURE{\epsfig{file= 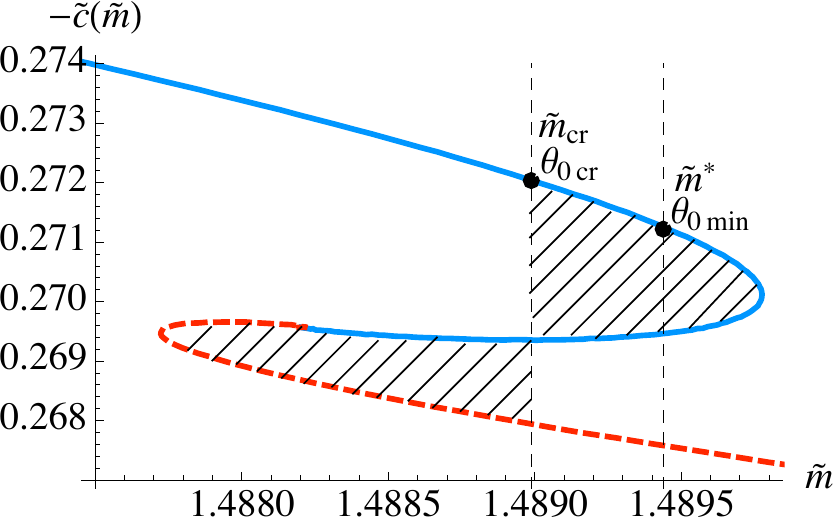,width=7.5cm} \epsfig{file= 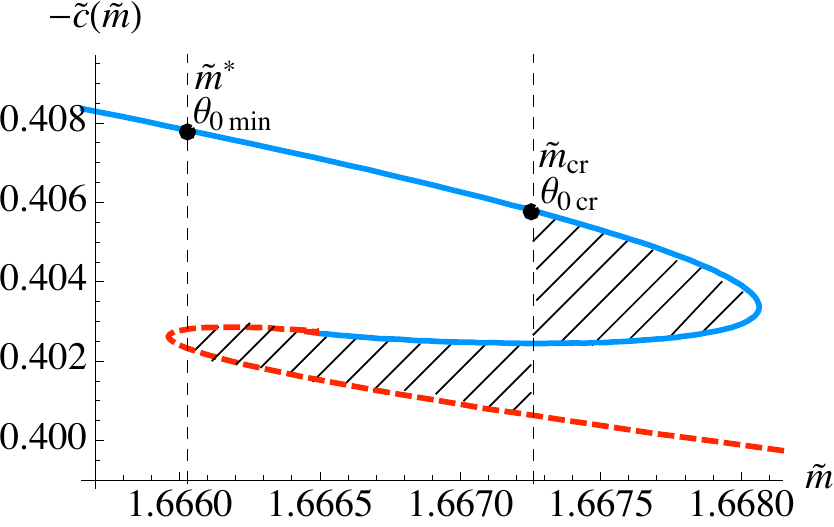,width=7.5cm}
\caption{\small  The physical scenarios described in the text for (left) ${\tilde E}<{\tilde E}_{\rm cr}$ and (right) ${\tilde E}>{\tilde E}_{\rm cr}$.  To the left, we have $\left(\theta_0\right)_{\rm cr}<(\theta_{0})_{\rm min}$, so the phase transition bypasses the singular branch of solutions. To the right, we have $\left(\theta_0\right)_{\rm cr}>(\theta_{0})_{\rm min}$, so the singular solutions appear on the resulting phase diagram. Further discussion is available in the text. }
\label{fig:scenarios}}
Summarizing the two possible scenarios mentioned above, we have:
\begin{enumerate}  
\item For sufficiently weak electric field or large temperature ($\tilde E< \tilde E_{\rm cr}$) the $(\theta_{0})_{\rm min}$ embedding is in the vicinity of the ``critical'' ($\theta_0=\pi/2$) embedding and is bypassed by the phase transition. The thermodynamically stable phases correspond to smooth Minkowski and black hole embeddings on the gravity side or meson gas and quark gluon plasma states in the dual gauge theory.
\item For $\tilde E>\tilde E_{\rm cr}$ the $(\theta_{0})_{\rm min}$ embedding is thermodynamically stable and corresponds to some bare quark mass $\tilde m^{*}<\tilde m_{\rm cr}$. This means that some of the classically stable black hole solutions, namely the one corresponding to bare quark mass in the range $\tilde m^{*}<{\tilde m}<\tilde m_{\rm cr}$, will have a conical singularity.
\end{enumerate}
There is an example of each such scenario in
figure~\ref{fig:scenarios}.
  \FIGURE{\epsfig{file= 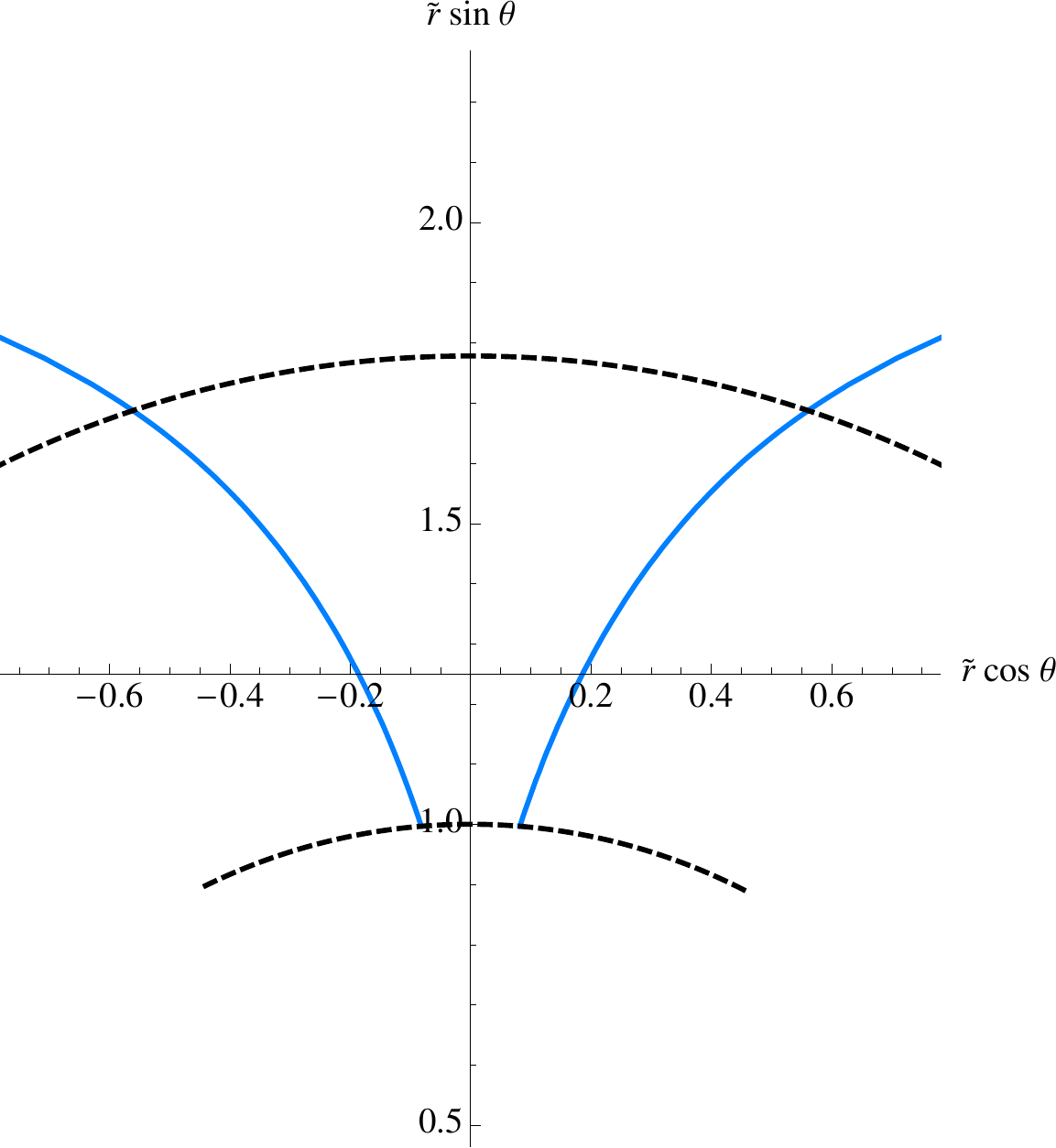,width=2in}  \epsfig{file= 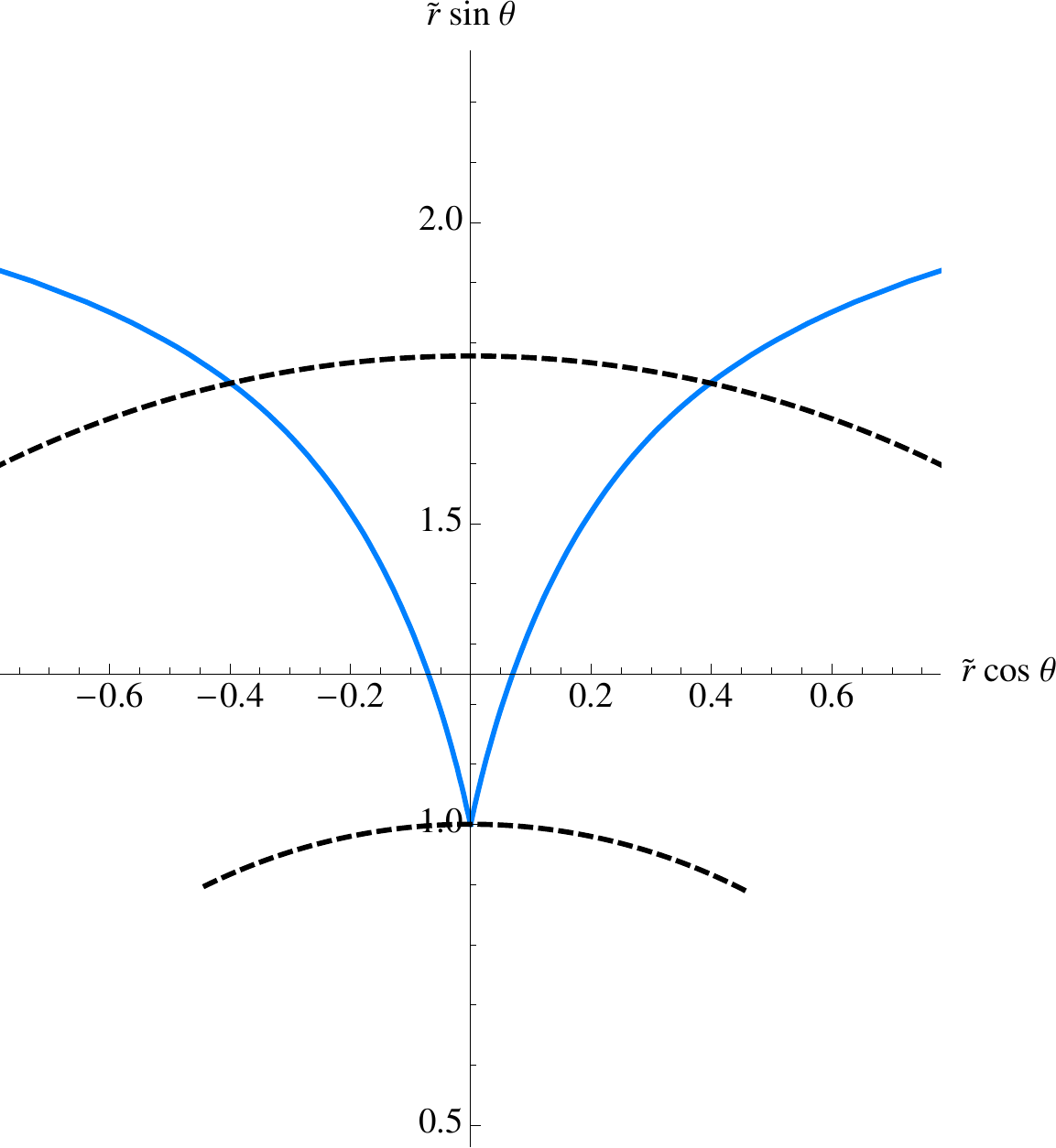,width=2in} \epsfig{file= 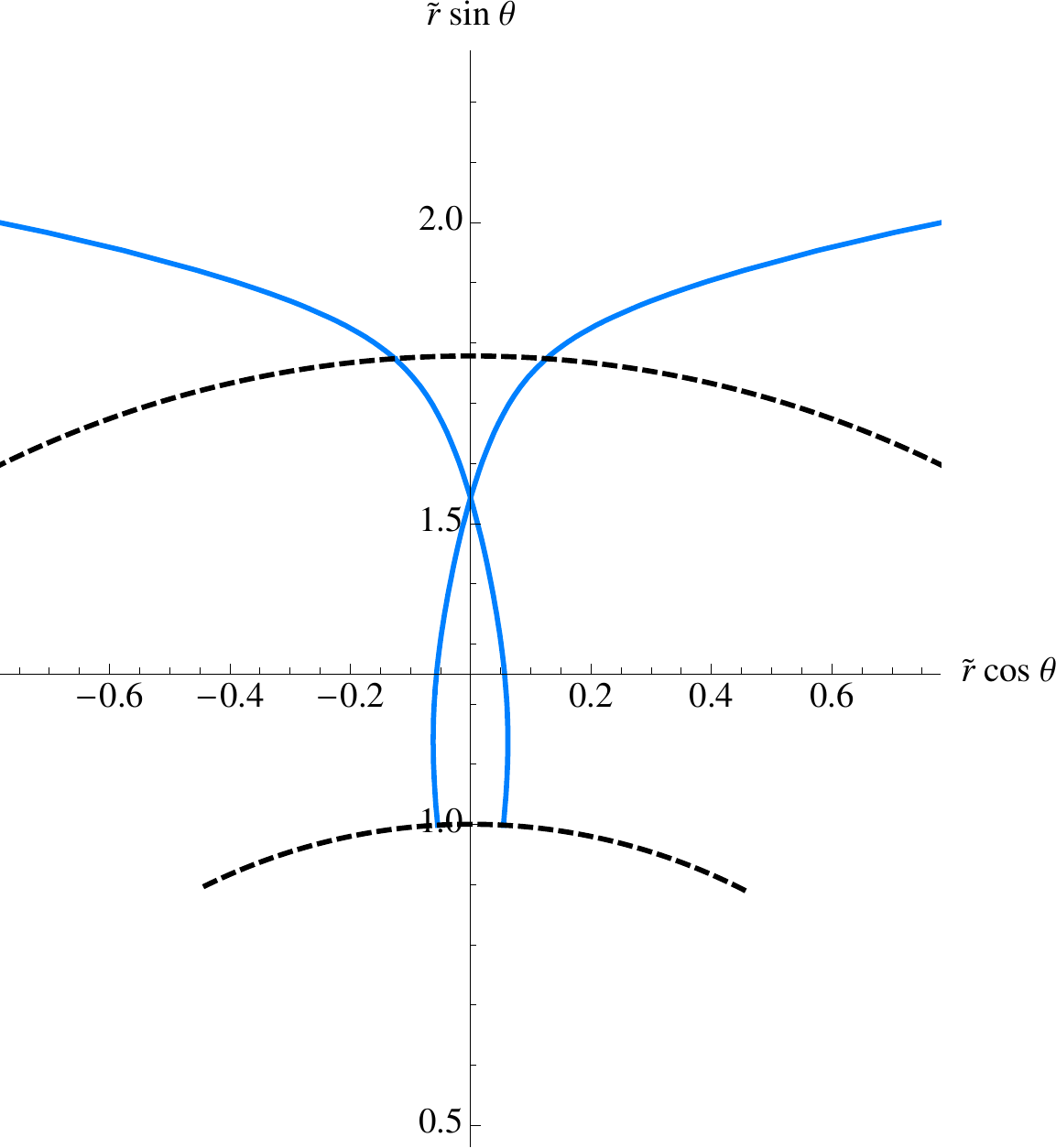,width=2in}
 \caption{\small  Examples of D7--brane embeddings for $\theta_0>(\theta_{0})_{\rm min}$, $\theta_0=(\theta_{0})_{\rm min}$ and $\theta_0<(\theta_{0})_{\rm min}$ respectively. The inner semi--circular segments represent the event horizon, while the outer semi--circular arcs correspond to the vanishing locus at $r=r_\ast$. The numerical value of the minimal angle is $\left(\theta_{0}\right)_{\rm min}\approx 1.345$ and the parameter $\tilde E=3$.}
\label{fig:B}}
This suggests a richer structure for the phase diagram than the one presented in figure~\ref{fig:phase} and in particular the existence of a special point at $\tilde E=\tilde E_{\rm cr}$ and $\tilde m_{\rm cr}(\tilde E_{\rm cr})$. The corresponding phase diagram is presented in figure \ref{fig:phase2}. The new choice of dimensionless variables on the axes is made in order to emphasize the existence of the non--smooth ({\it i.e.,} conically singular) black hole embeddings. The solid curve corresponds to the phase curve from figure~\ref{fig:phase} (but note that the axes are different here, and hence the shape) and the dashed curve separates embeddings with a conical singularity from smooth black hole embeddings (that lie below it). The area above the solid curve is in the stable meson phase and the area below the dashed curve is in the melted/dissociated phase.  The area between the curves corresponds to the embeddings that have a conical singularity before ending on the horizon.  The vertical dashed line corresponds to the critical value $\tilde E_{\rm cr}\approx 1.26$, below which the conical solutions are not part of the energetically favourable solutions and so do not complicate the story.  In order to relate the results in figure \ref{fig:phase2} to those of the zero temperature case, we first note that the $T=0$ dimensionless variable $\hat m$ can be related to the finite $T$ variable $\tilde m$ \emph{via}:
\begin{equation}
\hat m=\frac{m}{R\sqrt{E}}=\lim_{T\to0}\frac{\tilde m}{\sqrt{\tilde E}}\ .
\end{equation}
In addition, if $\hat{E} = b^2 \tilde{E} /R^2$ is to remain finite as $b \to 0$, $\tilde{E}$ must go to infinity.  Therefore, the $T \to 0$ limit corresponds to studying $\tilde{E} \to \infty$ in figure \ref{fig:phase2}.  We see that the phase curves from figure~\ref{fig:phase2} qualitatively approach the zero temperature (large electric field) case of the previous section, with the critical parameters $\hat m^{*}\approx 1.22$ and $\hat m_{\rm cr}\approx 1.32$.
\FIGURE{\epsfig{file= 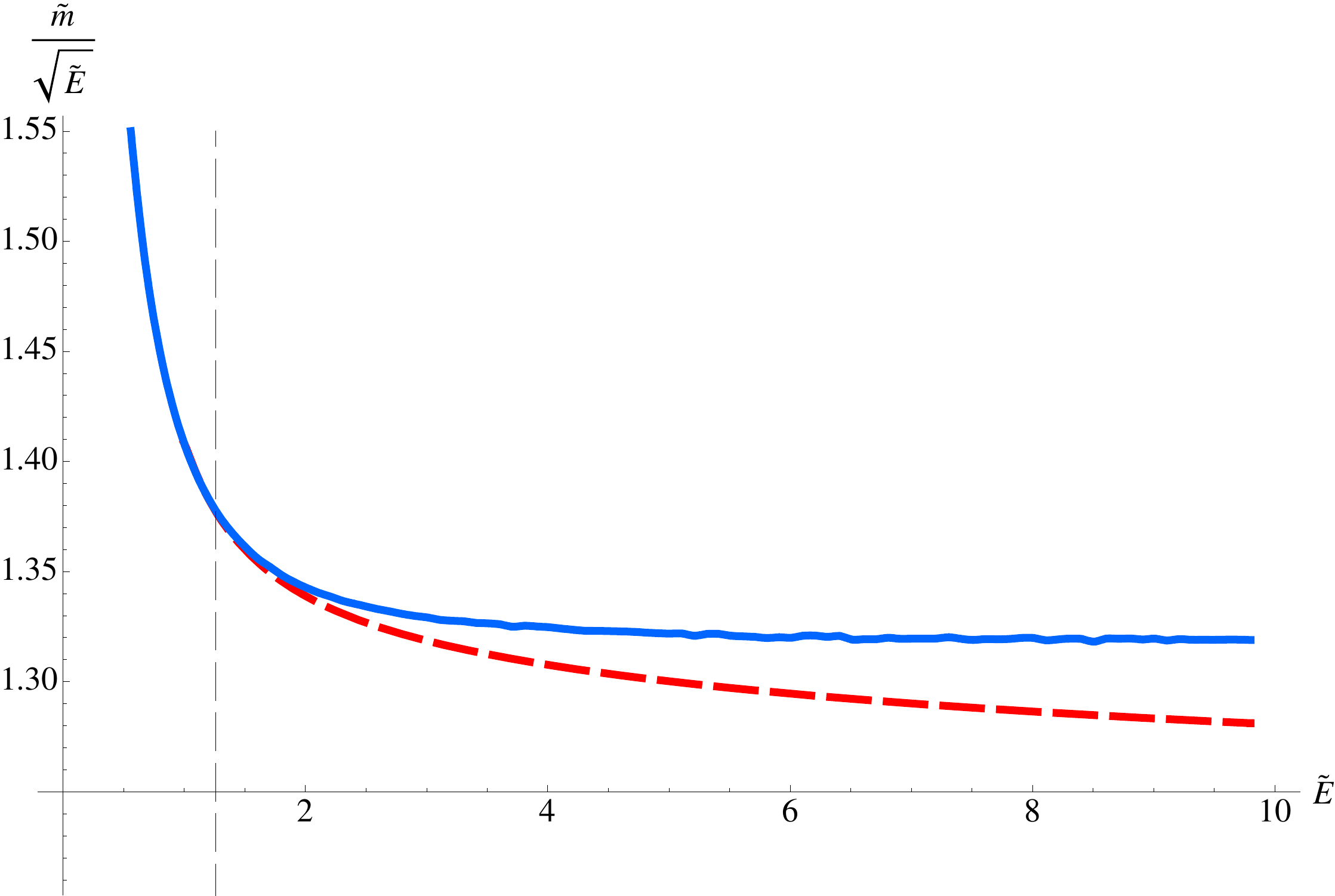,width=12cm}
\caption{\small The solid curve correspond to the phase curve from figure \ref{fig:phase} and the dashed curve separates embeddings with a conical singularity from smooth black-hole embeddings. The area above the solid curve is in the meson gas phase and the area below the dashed curve is in the quark--gluon plasma phase. The area between the solid and dashed curves  corresponds to embeddings which reach the black hole horizon  but have a conical singularity before reaching the event horizon. Their physical status remains to be determined but may be irrelevant for this particular phase diagram (see text).  The vertical dashed line corresponds to the critical value $\tilde E_{\rm cr}\approx 1.26$ where the conical solutions become thermodynamically relevant.} 
\label{fig:phase2}}
%
\section{Meson Spectroscopy} \label{section:meson}
We end by making some observations about the meson spectrum. To
proceed in studying the meson spectrum, we first absorb our gauge
field into the $B$ field to emphasize that it is of the same order in
$\alpha'$ at the classical level of our Dirac--Born--Infeld action.
This simplifies our notation since $A_\mu$ now denotes the second
order in $\alpha'$ gauge field.  To study the meson spectrum, we need
to consider quadratic (in $\alpha'$) fluctuations about our classical
embeddings for the D7--branes \cite{Karch:2002sh}.  To this goal, the
relevant pieces of the action are:
\begin{eqnarray} \label{eqt:action2}
S&=& - \frac{T_7}{g_s} \int d^8 \xi \sqrt{g_{ab} + B_{a b}+ 2 \pi \alpha' F_{a b}} + \left(2\pi \alpha'\right) \mu_7 \int_{\mathcal{M}_8} F_{(2)} \wedge B_{(2)} \wedge P\left[\tilde{C}_{(4)} \right]  \nonumber \\
&&  +\left(2\pi \alpha' \right)^2 \mu_7 \frac{1}{2} \int_{\mathcal{M}_8} F_{(2)} \wedge F_{(2)} \wedge P\left[C_{(4)}\right] \ ,
\end{eqnarray}
where
\begin{eqnarray*}
C_{(4)} &=& \frac{1}{g_s} \frac{u^4}{R^4} dt \wedge dx^1 \wedge dx^2 \wedge dx^3  \ , \\
\tilde{C}_{(4)} &=& -\frac{R^4}{g_s}  \left(1- \cos^4\theta \right) \sin\psi \cos\psi\ d \psi \wedge d \alpha \wedge d \beta \wedge d\phi  \ . \\
\end{eqnarray*}
To study the quadratic fluctuations, we consider an expansion for the transverse fields of the form:
\begin{eqnarray}
\theta &=& \theta_0(r) + 2\pi \alpha' \chi (\xi^a) \ , \\
\phi &=& 0 + 2 \pi \alpha' \Phi(\xi^a) \ ,
\end{eqnarray}
where $\theta_0$ is the classical embedding found in previous sections.  Expanding the action in equation \reef{eqt:action2} to second order in $2 \pi \alpha'$, we find the relevant terms for our fluctuation analysis are:
\begin{eqnarray}
- \mathcal{L}_{\chi^2} &=& \frac{1}{2} \sqrt{-E} E^{a b} G_{\theta \theta} \partial_a \chi \partial_b \chi -  \frac{1}{2} \sqrt{-E} E^{a b} \theta^{\prime 2} G_{\theta \theta}^2 E^{u u} \partial_a \chi \partial_b \chi \nonumber \\
&& + \frac{1}{2} \chi^2 \left[ \partial_\theta^2 \sqrt{-E} - \partial_u \left(  \theta' G_{\theta \theta} E^{u u} \partial_\theta \sqrt{-E}  \right) \right] \ , \\
- \mathcal{L}_{\Phi^2} &=& \frac{1}{2} \sqrt{-E} G_{\Phi \Phi} E^{a b} \partial_a \Phi \partial_b \Phi \ , \\
- \mathcal{L}_{F^2} &=& -\frac{1}{4} \sqrt{-E} S^{a b} S^{c d} F_{b c} F_{d a}  \ , \\
-\mathcal{L}_{F-\chi} &=& \frac{1}{2} \chi F_{a b} \left[ R^2 \partial_u \left( \sqrt{-E} \theta'  E^{u u} J^{a b} \right) -  J^{a b} \partial_\theta \sqrt{-E} \right] \ , \\
\mathcal{L}^{\mathrm{WZ}}_{F-\Phi} &=& -4 R^4 \cos^3 \theta \sin \theta \sin \psi \cos \psi B_{0 1}  F_{2 3} \Phi \ , \\
\mathcal{L}^{\mathrm{WZ}}_{F-F} &=& \frac{u^4}{8 R^4} F_{m n} F_{o p} \epsilon^{m n o p} \ .
\end{eqnarray}
Here $E_{a b} = g_{a b} + B_{a b}$, where $g_{a b}$ is the induced metric at the classical level.  $G$ will be reserved for the
space-time metric.  In addition, we use that $E^{a b} = S^{a b} + J^{a b}$, where $ S^{a b} = S^{ b a}$ and $J^{a b} = - J^{b a}$.  The indices $a, b, c, d$ range over the worldvolume coordinates, and the indices $m, n , o , p$ range over the directions $4,5,6,7$.  We have reverted to using the coordinate $u$ instead of $r$ since the equations are simpler in this coordinate.  The equations of motion derived from these Lagrangian terms allow us to consistently take $A_1(t,u) \neq 0$, $A_4(t,u) \neq 0$, and the remaining gauge fields to be zero.  This implies that the field $\chi$ is coupled to $A_1$ and $A_4$, whereas the field $\Phi$ decouples from the gauge fields.
Let us begin by studying the spectrum for the zero temperature case.  For simplicity, we consider fluctuations about the $\theta = 0$ embedding, in which case, the field $\chi$ decouples from the gauge field.  We consider an ansatz of the form:
\begin{equation} \label{eqt:ansatz_meson}
\chi \equiv \chi \left(\hat{u}, \hat{t} \right) = \xi(\hat{u}) {\rm e}^{- i \hat{\omega} \hat{t} } \ ,
\end{equation}
where $\hat{u} = u R \sqrt{\hat{E}} $, $\hat{t} = t \sqrt{\hat{E}}/ R$, and $\hat{\omega} = \omega R/ \sqrt{\hat E}$ are dimensionless quantities.  We would like to focus our attention to the equation of motion near the vanishing locus, in this case, at $\hat{u}=\hat{u}_\ast =1$.  The equation of motion has the form:
\begin{equation}
\xi''(\hat{u}) + \frac{1}{\hat{u}-\hat{u}_\ast } \left(1 + i \frac{\hat{\omega}}{\sqrt{6}} \right) \xi'(\hat{u}) + \frac{1}{\hat{u}-\hat{u}_\ast } \left( \frac{1}{2} - i  \frac{\hat{\omega}}{\sqrt{6}} + \frac{5 \hat{\omega}^2}{12} \right) \xi(\tilde{u}) = 0 \ .
\end{equation}
We emphasize that this equation assumes that $\tilde{T}$ is positive.  We find that this equation satisfies an ingoing wave solution of the form:
\begin{equation}
\xi (\hat{u}) = \left(\hat{u}-\hat{u}_\ast  \right)^{- i \hat{\omega} / \sqrt{6}} \zeta \left(\hat{u} \right) \ ,
\end{equation}
The outgoing wave solution corresponds to simply taking the complex conjugate of the ingoing wave solution:
\begin{equation}
\chi_{\rm out} = \chi^*_{\rm in} \ .
\end{equation}
In examples involving string worldsheets \cite{Gubser:2006nz,CasalderreySolana:2007qw}, such behavior was interpreted as an effective horizon for the worldsheet, and only the ingoing wave solution was considered physical.  Here, the ingoing solution implies that embeddings that touch the vanishing locus must fall towards the event horizon.  This is simply the statement that the conducting phase can only be associated with black hole embeddings.  We emphasize that the situation here is not exactly the same as in ref.~\cite{Gubser:2006nz,CasalderreySolana:2007qw}; the vanishing locus does not correspond to the horizon of the induced D7--brane metric.
However, there are indications that the gauge theory only sees the physics up to the vanishing locus.  The current $\tilde{T}$ is defined at $\tilde{u}_\ast$; the asymptotic behavior of the branes can be solved for by imposing initial conditions at $\tilde{u}_\ast$, so the bare quark mass and condensate can be identified without having to solve for the embedding up to the event horizon.  From our results above, the same holds true for the meson spectrum.  Therefore, one may be tempted to interpret the vanishing locus as an effective horizon; however, Ramond--Ramond (RR) charge conservation on the D7--brane requires us to continue the embedding up to the physical event horizon.  Perhaps for the questions we are asking, the gauge theory is ignorant of the physics inside the vanishing locus, and hence the conical singularities we have encountered and their resolution do not alter the phase diagrams we have presented.

Therefore, let us proceed and check the behavior of the fluctuations for the $T \neq 0$ case.  For a similar ansatz as before, near the vanishing locus, we again find ingoing wave solutions of the form:
\begin{equation}
\xi (\tilde{u}) = \left( \tilde{u} - \tilde{u}_\ast \right)^{- i a \tilde{\omega} } \zeta(\tilde{u}) \ ,
\end{equation}
where $\tilde{u} = u / b$, $\tilde{t} = b t / R^2$, $\tilde{\omega} = R^2 \omega / b$, and $a = \frac{\sqrt{4+10 \tilde{E}^2  + 6 \tilde{E}^4}}{2 \tilde{u}_\ast (2 + 3 \tilde{E}^2)}$.  $\tilde{u}_\ast$ determines the position of the vanishing locus and is given by $ \sqrt[4]{1+\tilde{E}^2}$.
Therefore, we find that the solution becomes exactly the incoming--wave quasinormal mode solution studied elsewhere \cite{Starinets:2002br,Hoyos:2006gb, Mateos:2006nu,Albash:2006ew}.  Therefore, this result confirms that we still have a ``melted'' phase for the mesons corresponding to black hole embeddings.  
%
\section{Conclusions}
It is very encouraging that non--trivial non--perturbative phenomena resulting from external fields such as those we have seen here (a dissociation phase transition, metal--insulator transition and the associated response current) can be so readily extracted in this kind of holographic study. We found that since the electric field works together with the presence of finite temperature, the resulting phase diagram which accounts for the effects of both is rather simple (ignoring the complication of the special conical solutions ---see below). This paper is a natural companion to our recent work \cite{Albash:2007bk} on the same system in an external magnetic field, where we mapped out somewhat richer phase structure, seeing how magnetically induced spontaneous chiral symmetry breaking (studied in this context in ref.~\cite{Filev:2007gb}) and the meson melting fit together in the diagram.

Clearly, the story is not quite complete in this electric case, since above a certain value of the electric field, some of our solutions develop conical singularities in the interior. We would like to know more about the nature of these solutions. As already stated, one\ possibility is that the solutions are locally (near the singularity) corrected by stringy physics, perhaps smoothing the conical points into throats that connect to the horizon. We would expect in that case that our phase diagram would largely remain intact, since the values of the masses and condensates for each solution are read off at infinity, far from where the conical singularity develops. The results for the relative free energies of those solution (compared to the other solutions at the same values of the mass) would be the same, and so the complete story would be unaffected.  Furthermore as noted in section \ref{section:meson}, key aspects of the meson spectrum lend support to the idea that  the phase diagram is insensitive to the local details of the conically singular features. It would be of value to find methods for making this indication more robust.

\acknowledgments
This work was supported by the US Department of Energy. We thank Johanna Erdmenger for conversations, and an anonymous referee for helpful comments that resulted in improved  presentation of some of the physics.

\providecommand{\href}[2]{#2}\begingroup\raggedright\endgroup
\bibliographystyle{JHEP}
\end{document}